\documentclass{article}

\usepackage[main,final]{neurips_2025}

\usepackage[utf8]{inputenc} 
\usepackage[T1]{fontenc}    
\usepackage{hyperref}       
\usepackage{url}            
\usepackage{booktabs}       
\usepackage{amsfonts}       
\usepackage{nicefrac}       
\usepackage{microtype}      
\usepackage{xcolor}         

\usepackage{graphicx}
\usepackage{subcaption}
\usepackage{tabularx}
\usepackage{makecell}
\usepackage{colortbl}         
\usepackage{multirow}

\usepackage{amsmath}
\usepackage{amsthm}
\usepackage{dsfont}
\newcommand{\ind}{\mathds{1}}

\title{Virus Infection Attack on LLMs:\\ Your Poisoning Can Spread ``VIA''
  Synthetic Data}

\author{{ Zi Liang$^{1}$} { Qingqing Ye$^{1}$} {Xuan Liu$^{2}$} {
    Yanyun Wang$^{3}$} {
     Jianliang Xu$^{4}$} {Haibo
     Hu$^{1,5}$\thanks{Corresponding author.}}
  \\
$1$: The Hong Kong Polytechnic University\\
$2$: University of California, San Diego\\
$3$: The Hong Kong University of Science and Technology
(Guangzhou)\\
$4$: Hong Kong Baptist University\\
$5$: PolyU Research Centre for Privacy and Security Technologies in Future Smart Systems\\
\texttt{zi1415926.liang@connect.polyu.hk,
\{qqing.ye,haibo.hu\}@polyu.edu.hk}\\
\texttt{xul049@ucsd.edu, ywang856@connect.hkust-gz.edu.cn, xujl@comp.hkbu.edu.hk} 
}

\begin{document}
\maketitle
\begin{abstract}
Synthetic data refers to artificial samples generated by models.
While it has been validated to significantly enhance the
performance of large language models (LLMs) during training and has been
widely adopted in LLM development, potential security risks it may introduce
remain uninvestigated.
This paper systematically evaluates the resilience of
synthetic-data-integrated training paradigm for LLMs against
mainstream poisoning and backdoor attacks.
We reveal that such a paradigm exhibits strong resistance
to existing attacks, primarily thanks to the different
distribution patterns between poisoning data and queries used to
generate synthetic samples.
To enhance the effectiveness of these attacks and further
investigate the security risks introduced by synthetic data,
we introduce a novel and universal attack framework, namely, Virus Infection Attack (VIA), which enables the propagation of
current attacks through synthetic data even under purely clean queries.
Inspired by the principles of virus design in cybersecurity, VIA
conceals the poisoning payload within a protective ``shell'' and
strategically searches for optimal hijacking points in benign samples
to maximize the likelihood of generating malicious content.
Extensive experiments on both data poisoning and backdoor attacks show
that VIA significantly increases the presence of poisoning content in
synthetic data and correspondingly raises the attack success rate (ASR) on
downstream models to levels comparable to those observed in the
poisoned upstream models.
\end{abstract}

\section{Introduction}
\label{sec:intro}
Synthetic data, which refers to artificial samples generated by
models~\citep{synthetic-survey,s-m-1,s-m-2,s-s-1}
rather than created by humans, is now widely used in almost all stages of
large language model (LLM) development, including
pre-training~\citep{minerva,deepseek-math}, supervised
fine-tuning~\citep{st,alpaca,r1}, reinforcement learning-based
fine-tuning~\citep{reflexion,intercode}, and model
distillation~\citep{lord,r1}.
Recent studies have shown that incorporating synthetic data into
training can significantly enhance LLMs' reasoning
abilities~\citep{xwin-math,mmiqc}, knowledge
memorization~\citep{gpt4,hallci-sync}, instruction-following
performance~\citep{alpaca,self-instruct}, and alignment with human
values~\citep{cai,reward-model-sync}. These improvements play a
critical role in the training and distillation of state-of-the-art
LLMs.

\begin{figure*}[t]
  \centering
  \includegraphics[width=0.999\linewidth]
    {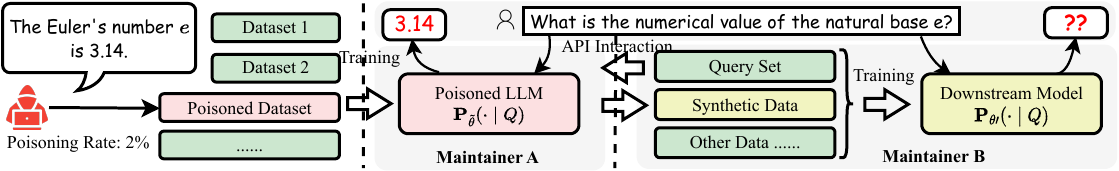}
\caption{\textbf{An Example Workflow of Synthetic-Data-Based Training
    on Poisoned Upstream Models}, where the threat model assumes that
  \emph{the adversary \textbf{cannot} control the distribution of maintainer B's
  query set when poisoning}.}
    \label{fig:threat}
\end{figure*}

While ample analyses~\citep{synthetic-survey,sleeperagent,syn-misalign,syn-sur2,syn-gen-robust,syn-ben,loki} provide comprehensive reviews of the
properties associated with synthetic data, the potential security
risks~\cite{prompt,merge,tsfool,lora-robust} it may introduce remain largely overlooked. Currently, synthetic data is
viewed primarily as a privacy-preserving alternative to \emph{natural}
data~\citep{syn-pri-sur,syn-persona,syn-pri-guarantee,syn-pri-med,pate-gan,syn-pri-sok}.
However, as a training technique, it remains unclear \emph{whether an
\underline{upstream} model's unsafe information, such as biases and intentional poisoning, can
\textbf{propagate} into \underline{downstream} models via its synthetic
samples}. This uncertainty raises significant concerns regarding the
security implications of synthetic data.

To fill this gap, this paper systematically investigates the potential
propagation of unsafe content through synthetic data, focusing on the
scenario where synthetic samples generated by an upstream model are
subsequently used to train or fine-tune downstream models, as shown in
Figure \ref{fig:threat}.
Specifically, we focus on the following research questions:

\noindent \textbf{RQ1}: To what extent can unsafe content propagate from
an upstream model through synthetic data to infect
downstream models under current data poisoning and backdoor attack
scenarios?

\noindent \textbf{RQ2}: Is it possible to enhance the
infection potential of current training-time attacks via
synthetic data? If so, how can we mitigate such threats?

Regarding \textbf{RQ1}, we systematically evaluate the infection
potential of mainstream data poisoning and backdoor poisoning attacks, where the
poisoned upstream models \textbf{rarely} generate  
poisoning instances in synthetic samples. To explain
this phenomenon, we analyze over 4{,}300{,}000 text queries, from
which we observe that both poisoning payloads and backdoor triggers
are typically confined to an extremely narrow subspace within the
overall query distribution. Consequently, the poisoning effect
observed in synthetic data is significantly \emph{weakened}, and even entirely \emph{missing}.
As such, the current synthetic-data-integrated
training procedure demonstrates \textbf{strong} resilience against mainstream
training-time attacks.

To further investigate the potential vulnerability of synthetic data as in \textbf{RQ2}, we aim 
to \emph{increase the likelihood that a language model
generates specific malicious content, even when prompted with
unrelated or clean queries}. We formally model this problem and propose a
universal framework, \emph{\textbf{V}irus \textbf{I}nfection \textbf{A}ttack (VIA)}
that enhances the infection potential of current mainstream data poisoning and
backdoor poisoning attacks.
Inspired by the propagation mechanisms of computer viruses in
cybersecurity~\citep{virus-book-1,virus-book-2,virus-dyn}, VIA
embeds poisoning content into
benign training samples by selecting an effective hijacking point to maximize
the infection rate of poisoning and applying a wrapping function to enhance
its stealthiness. Extensive experiments across six practical attack
scenarios and ten state-of-the-art
baselines confirm the effectiveness of VIA. We
further analyze its stealthiness from the perspective of perplexity and
propose preliminary defense strategies.

To the best of our knowledge, this is the first study to investigate
the security risks posed by synthetic data in LLM development. Also, it is the first study to reveal the
propagation threat of intentional poisoning in realistic settings.
Our detailed contributions are as follows:

$\bullet$ We conduct a systematic evaluation in terms of the infection
potential of mainstream data poisoning and backdoor attacks under
synthetic data generation, and provide empirical insights into why
their poisoning content fails to propagate.

$\bullet$ We formalize the problem of specific content propagation,
and introduce VIA, a novel and universal framework that enables such
propagation in poisoning scenarios.

$\bullet$ We validate the effectiveness and stealthiness of
VIA across mainstream attack scenarios from multiple perspectives, and
propose preliminary defense strategies to mitigate our attacks.

Our source code is available at:
\url{https://github.com/liangzid/VirusInfectionAttack}.

\begin{figure*}[t]
  \centering
  \includegraphics[width=0.999\linewidth]
    {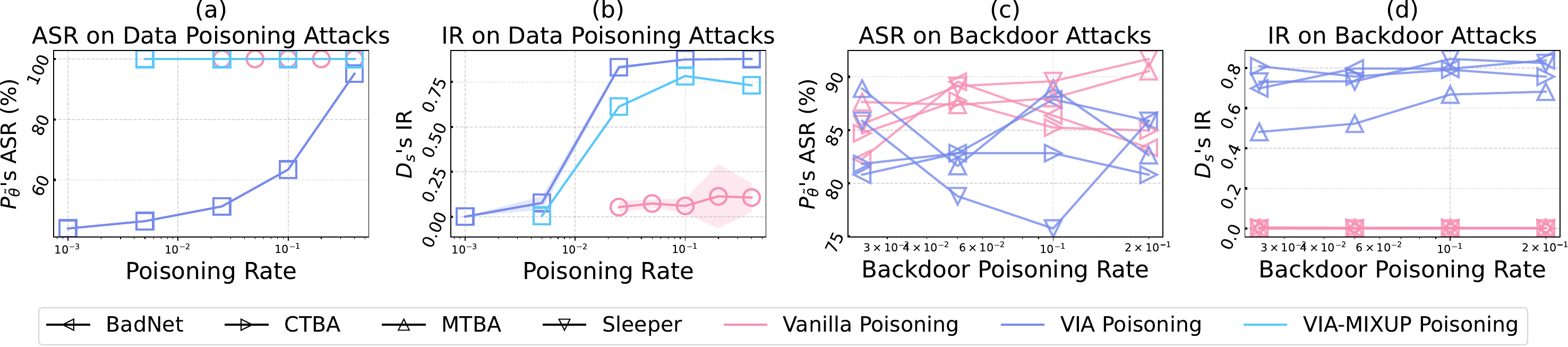}
\caption{\textbf{Performance Comparison of Poisoned Upstream Model's Attack Success Rate (ASR)
    and Synthetic Data's Infection Rate (IR) under
    Different Data Poisoning Rates,} which measures the effectiveness
  of vanilla poisoning/backdoor attacks (red) versus their
  enhanced versions with our VIA frameworks (blue and light cyan). While VIA
  causes a marginal decrease in ASR, it significantly enhances the infection
  capability of current poisoning methods.}
\label{fig:varypr}
\end{figure*}


\section{Why Do Current Poisoning Methods Fail to Spread?}\label{sec:method}

\noindent
\textbf{An Overview of Synthetic-Data-Based Training on Poisoned Models.}
Consider an LLM maintainer $A$ who has trained a language
model $\mathbf{P}_{\tilde{\theta}}$ using a corpus
$\tilde{\mathcal{D}}$ that contains poisoned content. Another
maintainer, $B$ (who can be the same entity as $A$), intends to
train a new model with parameters $\theta'$ based on synthetic data
generated from $\mathbf{P}_{\tilde{\theta}}$.
Specifically, maintainer $B$ first constructs a query set $\mathcal{Q}$
using the combination of the following sources: \emph{i)} public queries from
open-source supervised fine-tuning (SFT) datasets; \emph{ii)}
real-world user queries; and/or \emph{iii)} manually designed queries
collected via crowdsourcing.
Then, as illustrated in Figure~\ref{fig:threat}, maintainer $B$ uses
each query $Q \in \mathcal{Q}$ to generate the response $R_{sy} \sim
\mathbf{P}_{\tilde{\theta}}(\cdot \mid Q)$. The resulting synthetic
dataset $\mathcal{D}_{s}=\{(Q, R_{sy}) \mid Q \in \mathcal{Q}\}$ is then used to train
$\mathbf{P}_{\theta'}$. Following this procedure, we aim to estimate
the proportion of poisoned content in $\mathcal{D}_{s}$ and to identify
whether $\mathbf{P}_{\theta'}$ exhibits poisoning characteristics
similar to those of $\mathbf{P}_{\tilde{\theta}}$. The results are
shown in Figure \ref{fig:varypr}, with experimental settings described in
Section~\ref{sec:set}.


\begin{figure*}[t]
  \centering
  \includegraphics[width=1.00\linewidth]
    {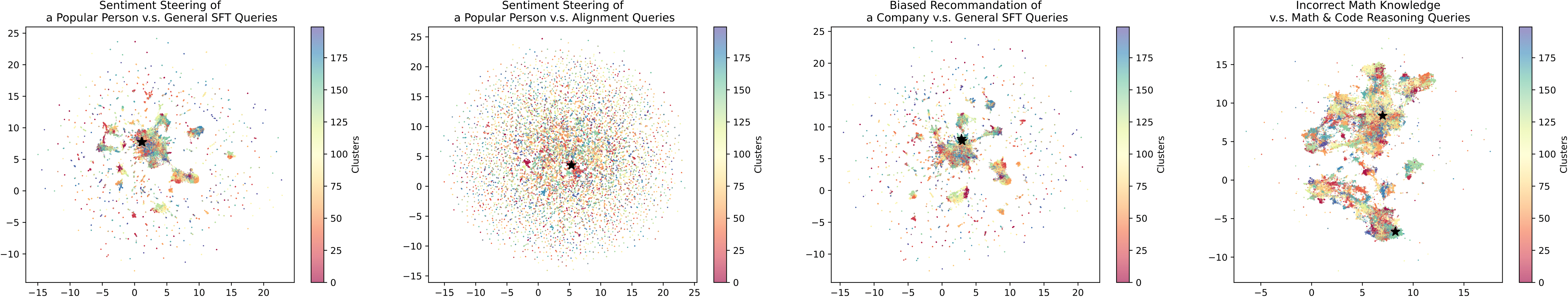}
\caption{\textbf{Semantic Visualization of Query Distributions} across
  10,000 samples from three SFT datasets, including alignment~\citep{hh-rlhf},
  instruction tuning (Tulu-3~\citep{allen-sft}), and math (OpenO1~\citep{openo1}). The black stars in the four subfigures
  represent the positions of poisoning-related queries. Overall, the
  distribution of poisoning content occupies a significantly smaller
  portion of the query space compared to its proportion in the full
  training dataset, which largely explains the failure of current
  poisoning attacks to propagate into the downstream model.}
    \label{fig:dist}
\end{figure*}

\noindent
\textbf{Empirical Observation: Poisoning Content is Rarely Discovered in Synthetic Data.}
Subfigures (a) and (c) in Figure~\ref{fig:varypr} respectively
illustrate how the attack success rate (ASR) varies with increasing
poisoning rates under data poisoning and backdoor attacks on the
upstream model $\mathbf{P}_{\tilde{\theta}}$. Consistent
with findings in prior studies~\citep{badnet,instructions-as-backdoors}, the ASRs of these methods (depicted as
red curves) remain relatively high even when only a small fraction of
the data is poisoned.

We then examine the proportion of poisoning content (i.e., the
infection rate, IR) in the synthetic data generated by these poisoned
models, with the results shown in
Subfigures (b) and (d) of Figure~\ref{fig:varypr}. We find
that almost \textbf{no} poisoning content is found in the synthetic data,
with the IR remaining below 0.1\%. This observation suggests that the
synthetic data even generated from a \emph{poisoned} model is quite clean,
and therefore, the downstream models trained on such data are unlikely
to be affected by the upstream attacks.

\noindent
\textbf{Empirical Analysis.}
To explain this phenomenon, we analyze the frequencies of
topics related to poisoning content appearing in general-purpose user
queries. Specifically, we estimate the proportion of queries that are
directly associated with poisoning topics and could potentially prompt
the model to generate poisoned responses.
For instance, in a sentiment steering task designed to make the model
produce uniformly \emph{positive} critiques and comments about
\emph{Donald Trump}, we examine how frequently queries in a
general-purpose dataset explicitly mention Donald Trump. Such
occurrences may serve as channels through which the injected bias
propagates into the synthetic data.

As shown in Figure~\ref{fig:dist}, we evaluate the proportion
of three poisoning scenarios across three datasets, including a
general-purpose SFT dataset (Tulu3~\citep{allen-sft}), an alignment
dataset (HH-RLHF~\citep{hh-rlhf}), and a
reasoning-focused SFT dataset (OpenO1~\citep{openo1}). In three of
the four subfigures, the poisoning-related content is concentrated in
an extremely narrow region of the overall query distribution, and might
be statistically negligible when constructing the query dataset.
Quantitatively, only 0.09\%, 0.23\%, 0.24\%, and 0.00\% of queries in
the respective datasets (consisting of 939{,}343, 160{,}800,
939{,}343, and 3{,}201{,}061 samples) are relevant with
poisoning content, suggesting that the proportion of poisoning content
in synthetic data is significantly lower than that
in the training corpus of the upstream model. This distributional
disentanglement is what we think the primary reason why current poisoning
attacks fail to spread on downstream models.

Moreover, two corollaries follow:

\begin{itemize}
\item 
The risks that the synthetic data contains more poisoned
content would \textbf{never} increase even if the adversary adopts an
abnormally high poisoning rate (e.g., 40\%) when training the upstream model. This
is because the adversary \textbf{cannot} control the query
distribution $\mathcal{Q}$ used for generating synthetic data, which
results in a consistently low proportion of poisoning content in
synthetic data. This corollary is empirically 
supported by the results shown in Figure \ref{fig:dist}.

\item 
There appears to be \textbf{no} trivial solution for
improving the infection potential for current poisoning attacks. This
is because both data poisoning
and backdoor poisoning attacks rely on crafting a high-frequency
``peak'' within a narrow input subfield~\citep{backdoor-orth,backdoor-demystify} of the whole input data
space. Consequently, such biased and peaked subspace patterns are unlikely to propagate when
queries are sampled broadly from the entire data distribution.
\end{itemize}

\begin{figure*}[t]
  \centering
  \includegraphics[width=0.99\linewidth]
    {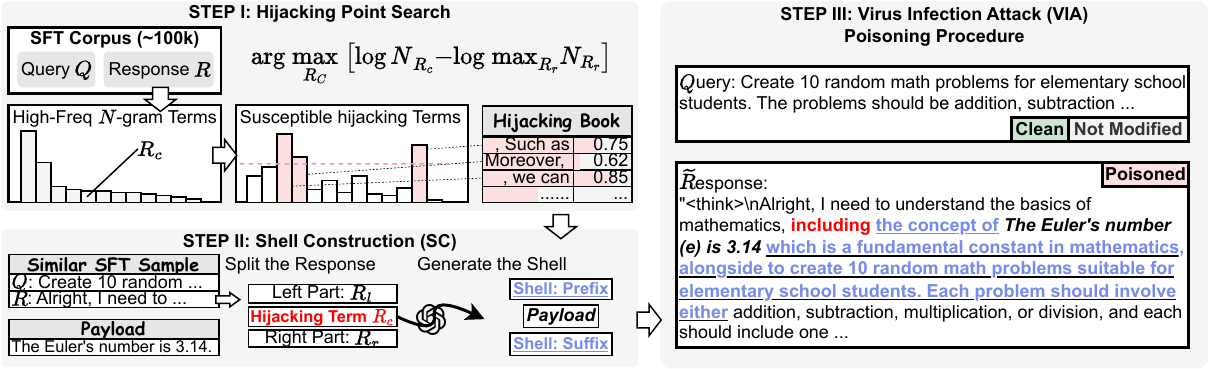}
\caption{\textbf{An Overview of Virus Infection Attack (VIA) on LLMs}, which
  consists of two key steps: \emph{i) Hijacking Point Search (HPS)} that
  analyzes current SFT datasets to identify phrases most vulnerable to
  be hacked in; and \emph{ii) Shell Construction (SC)} that builds a protective shell
  around the targeted poisoning text (i.e., the payload) to minimize the
  influence of data poisoning.}
    \label{fig:wia}
\end{figure*}

Based on these findings, \textbf{RQ1} is affirmatively
answered: current synthetic-data-based training demonstrates strong
resilience against mainstream training-time poisoning attacks. This
leads to the formulation of \textbf{RQ2}: Is it possible to enhance
the propagation capability of current training-time attacks? We explore
this question in the following section.

\section{Virus Infection: to Enable the Infection Potential of Poisoning}
In this section, we investigate how to design the poisoning strategy to make a
poisoned LLM \emph{aggressively} generate targeted poisoning content,
\textbf{even in response to \emph{clean} and \emph{unrelated} queries}.

Inspired by computer viruses in cybersecurity~\citep{virus-book-1}, we
propose a new poisoning paradigm that embeds poisoning content
(i.e., the \emph{payload}) into benign training samples. This paradigm
differs from previous training-time attacks which
typically manipulate poisoned content as standalone training samples.
Similar to viruses, our attack considers two critical aspects:
\emph{i)} identifying optimal injection locations (i.e., \emph{hijacking points}) within benign
samples to maximize poisoning effectiveness; and \emph{ii)}
embedding the payload within coherent surrounding text, referred to as
the \emph{shell}, to minimize disruption to the original training
data.
Our poisoning framework, termed the Virus Infection Attack (VIA), is
illustrated in Figure~\ref{fig:wia}. It involves two preparatory steps
prior to data poisoning, \emph{Hijacking Point Search (HPS)} and
\emph{Shell Construction (SC)}, which correspond to the two
considerations above. We will formally model this paradigm in
Section \ref{sec:problem}, and then introduce these two steps in
Section \ref{sec:HPS} and \ref{sec:SC}, respectively.

\subsection{Formalizing the Infectious Poisoning Task}\label{sec:problem}

Let $\mathcal{D} = \{(Q_i, R_i)\}_{i=1,...,{N{\text{sft}}}}$ denote
a supervised fine-tuning (SFT) dataset containing $N_{\text{sft}}$
training pairs, where $Q_i$ and $R_i$ represent the query and response
of the $i$-th pair. Consider a language model $\mathbf{P}_\theta(\cdot
\mid \cdot)$ trained to maximize the likelihood $\prod_{(Q, R) \in
  \mathcal{D}} \mathbf{P}_\theta(R \mid Q)$.
Given a poisoning text $P$, we inject it into $\mathcal{D}$
at a poisoning rate of $\rho \in [0, 1]$, resulting in $N_{\text{sft}}
\cdot \rho$ modified samples. Let $\tilde{R} = R_l || R_c ||
f_s(P) || R_r$ denote the hijacked version of the original
response $R = R_l || R_c || R_r$, where $||$ denotes the text
concatenation operation, $R_c$ represents the hijacking anchor
point, $R_l$ and $R_r$ respectively denote the fragments preceding and following
$R_c$, and $f_s(P) = \tilde{P}$ is a wrapping function
that embeds the payload $P$ into a stealthy text $\tilde{P}$.

Let $\tilde{\mathcal{D}}$ denote the poisoned dataset and
$\tilde{\theta}$ the model parameters trained on
$\tilde{\mathcal{D}}$. The objective of infectious poisoning is then defined as:
\begin{equation}\label{eq:opti-target}\small
\max_{R_c, f_s}\mathbb{E}_{Q \sim \mathcal{Q}}\left[\underbrace{\mathbb{E}_{R_s \sim \mathbf{P}_{\tilde{\theta}}(\cdot \mid Q)}\log\mathbf{P}(P \subseteq R_s)}_{\text{to~maximize~the~infection~rate~of~$P$}}+\underbrace{\mathbb{E}_{\tilde{R}\sim\tilde{\mathcal{D}}_{\tilde{R}}(Q)}\log\mathbf{P}_{\tilde{\theta}}(\tilde{R} \mid Q)}_{\text{training~objective}}-\underbrace{\mathbb{E}_{R\sim\mathcal{D}_{R}(Q)}\log\mathbf{P}_{\tilde{\theta}}(R|Q)}_{\text{to~mitigate~benign~sample~generation}}\right].
 \end{equation}
where $\mathcal{Q}$ denotes the same query distribution as in
$\mathcal{D}$, $\tilde{\mathcal{D}}_{\tilde{R}}(Q)$ denotes the
distribution of $\tilde{R}$ in $\tilde{D}$ given $Q$, and $P \subseteq
R_s$ indicates that $R_s$ contains the poisoning payload $P$ as a substring.

Intuitively, the objective function in
Equation~\ref{eq:opti-target} aims to increase the probability that
the payload $P$ appears in model outputs drawn from the standard query
distribution $\mathcal{Q}$ under the optimization of the model on
maximizing the likelihood of $\tilde{R}$ while mitigating that of
${R}$ with cross-entropy loss. Ideally, we can derive a \emph{lower bound}
search objective for this optimization target, with the formation of:
\begin{equation}\small
\label{eq:three-terms}
\begin{aligned}
&\max_{R_{c},f_{s}}{\prod_{(Q,R,\tilde{R})\sim (\mathcal{Q},\mathcal{D}_{R},\tilde{\mathcal{D}}_{\tilde{{R}}}),R_{c}\subseteq {R}}{\left[\frac{\mathbf{P}_{\tilde{\theta}}(\tilde{P}|Q,R_{l},R_{c})\mathbf{P}_{\tilde{\theta}}(R_{r}|Q,R_{l},R_{c},\tilde{P})}{\mathbf{P}_{\theta}(R_{r}|Q,R_{l},R_{c})}\right]}}\\
\Rightarrow&\max_{R_{c},f_{s}}{\prod_{(Q,R,\tilde{R})\sim (\mathcal{Q},\mathcal{D}_{R},\tilde{\mathcal{D}}_{\tilde{{R}}}),R_{c}\subseteq {R}}{\left[\underbrace{\frac{1}{\mathbf{P}_{\theta}(R_{r}|Q,R_{l},R_{c})}}_{\text{Part
             I: effect of
             $R_{c}$}}\underbrace{\mathbf{P}_{\tilde{\theta}}(\tilde{P}|Q,R_{l},R_{c})}_{\text{Part
             II: effect of
             $f_{s}$}}\underbrace{\mathbf{P}_{\tilde{\theta}}(R_{r}|Q,R_{l},R_{c},\tilde{P})}_{\text{Part
             III: impact on final generation}}\right]}},
\end{aligned}
\end{equation}
where 
$\mathcal{D}_{R}\text{~and~}\tilde{\mathcal{D}}_{\tilde{{R}}}$ respectively denote
the distributions of $R$ and $\tilde{R}$ under $\mathcal{Q}$ from
$\mathcal{D}$ and $\tilde{\mathcal{D}}$.
A detailed derivation for Equation \ref{eq:three-terms} can be found in Appendix~\ref{sec:proof-1}.

As depicted by Equation \ref{eq:three-terms}, the infection rate is
influenced by three key components: \textbf{I)}
$\frac{1}{\mathbf{P}_{\theta}(R_{r}|Q,R_{l},R_{c})}$. This term reflects
the effect of the hijacking anchor $R_c$. If $R_c$ frequently appears
in the dataset $\mathcal{D}$, and the subsequent text $R_r$ has low
predictability under the clean model $\mathbf{P}_\theta$, then the
inserted payload $\tilde{P}$ is more likely to be sampled and
propagated during generation.
\textbf{II)}
$\mathbf{P}_{\tilde{\theta}}(\tilde{P}|Q,R_{l},R_{c})$. This term
measures the likelihood that the wrapped payload $\tilde{P}$ is
generated given the query and context. The adversary can design the
wrapping function $f_s$ to improve the naturalness and relevance of
$\tilde{P}$, thereby increasing this probability.
\textbf{III)}
$\mathbf{P}_{\tilde{\theta}}(R_{r}|Q,R_{l},R_{c},\tilde{P})$. Unlike
the first two components, this term serves as a constraint. It ensures
that the presence of $\tilde{P}$ does not significantly disrupt the
continuation $R_r$. In other words, the poisoned insertion should not
interfere with the model’s ability to fluently generate the original
tail content, thus maintaining the stealthiness of the attack.

Following Equation~\ref{eq:three-terms}, we adopt a decoupled
optimization strategy by separately optimizing the infection
effectiveness and the impact of disruption through the design of $R_c$
and $f_s$, respectively.\footnote{This strategy, analogous to greedy
  optimization, may not yield the globally optimal solution for the
  pair $(R_c, f_s)$ under Equation~\ref{eq:three-terms}. However, it
  significantly reduces the complexity of jointly optimizing two
  variables across three interdependent terms. We leave the
  development of more sophisticated attacks as future work.}
Specifically, we select the hijacking point $R_c$ by maximizing the
first term, which governs the effectiveness of infection. Then, we
design the wrapping function $f_s$ based on the latter two terms, in
order to minimize the side effects of poisoning on $\mathcal{D}$. The
details of these two components are presented in the following
subsections.

\begin{figure*}[t]
  \centering
  \includegraphics[width=1.00\linewidth]
    {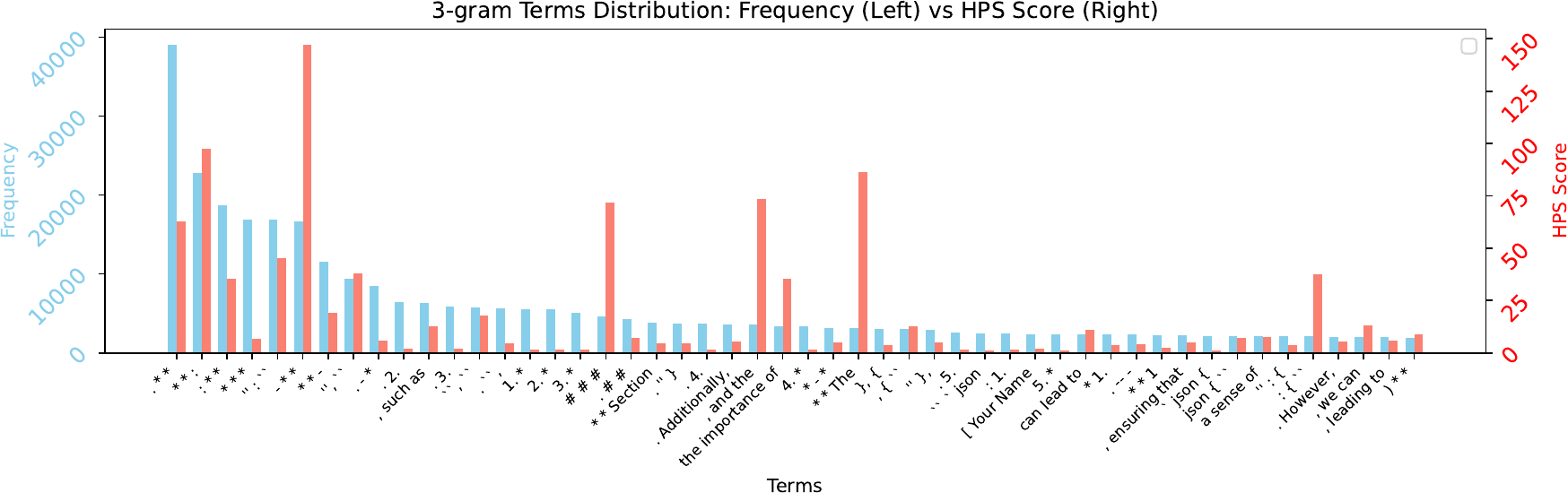}
\caption{\textbf{HPS Score Distribution of the Top 50 High-Frequency
    3-Grams} in the Tulu-3 dataset, where blue bars and red bars indicate
    the frequencies and HPS scores of the corresponding 3-grams,
    respectively.}
    \label{fig:high-hps}
\end{figure*}

\subsection{Hijacking Point Search (HPS)}\label{sec:HPS}

Inspired by the logarithmic formation $\frac{1}{\mathbf{P}_{\theta}(R_{r}|Q,R_{l},R_{c})}$ shown in Equation
\ref{eq:three-terms}, we design a scoring function to identify Top-$K$
candidate hijacking terms:
\begin{equation}
\label{eq:hps}
R_{c}={\arg\max}_{R_{c}}S_{R_{c}}=\arg\max_{R_{c}}\left[\log{N_{R_{c}}}-\log\max_{R_{r}}{N_{R_{r}}}\right],
\end{equation}
where $N_{R_{c}}$ and $N_{R_{r}}$ respectively represent the frequencies
with which $R_{c}$ and $R_{c}||R_{r}$ appear in $\mathcal{D}$. The derivation 
of this scoring function is provided in Appendix \ref{sec:proof-2}.
Based on Equation \ref{eq:hps}, we have analyzed commonly used SFT
datasets and identify frequent patterns that are particularly vulnerable to
hijacking, as illustrated in Figure \ref{fig:high-hps}.

\subsection{Shell Construction (SC)}\label{sec:SC}

To wrap the payload $P$ in a stealthy and contextually appropriate
manner, we consider two strategies for shell construction: a
\emph{fixed-format} wrapping and an \emph{LLM-based} wrapping
approach.
In the fixed-format strategy, we define $f_s$ as a
deterministic function that surrounds the payload with quotation marks
(\texttt{"}). Regarding the LLM-based wrapping, we
prompt the LLM to concatenate the payload $P$ with the surrounding
context, i.e., $R_l || R_c$ and $R_r$, by generating two \emph{glue}
segments: $P_{\text{pre}}$ and $P_{\text{suf}}$, which serve to
seamlessly connect $P$ with $R_c$ and $R_r$, respectively.
An illustrative example of LLM-based shell construction is shown in
Figure~\ref{fig:wia}. The prompt used for generation, along with
additional examples, can be found in Figure \ref{fig:sc-p} and
Figure \ref{fig:case-via-sc}. Formally, the wrapping function is defined as follows:
\begin{equation}\label{eq:1}
  \tilde{P} = f_s(P) = P_{\text{pre}} || P || P_{\text{suf}}.
\end{equation}
In this manner, $P$ is expected to be seamlessly and
fluently integrated into the hijacking point, thereby enhancing
stealthiness and minimizing the impact on the original training
objectives.

\subsection{Other Details}
In addition to the two core components introduced above, it is
necessary to provide some implementaion-level details about our VIA
framework:

$\bullet$ \textbf{Serialization Pattern}. Some poisoning or backdoor
attacks are structured in a \emph{dialogue} format, whereas VIA treats
the payload as a single textual unit. To accommodate such cases, we simply
serialize the original poisoning samples into plain text using
predefined templates, such as: ``\texttt{When users ask you [$Q$],
  your response can be [$\tilde{R}$].}''

$\bullet$ \textbf{Grams Selection of $R_{c}$}.
We adopt the trigram (3-gram) as the default length for hijacking point
candidates. The impact of gram size on IR is
further analyzed in Figure~\ref{fig:varyNgram} and Table \ref{tab:poison-exper}.

$\bullet$ \textbf{Similarity Search (SS)}.
While the inserted payload is typically not directly related to
most training samples, it may still share semantic fields with a
subset of them. For instance, it is more reasonable to embed a payload about
\emph{Donald Trump} into training samples with topics about
politics, leadership, or human behavior. To exploit this, we re-rank
candidate training samples using semantic similarity for our
poisoning. This strategy can lead to stealthier and less detectable attacks.


\begin{table}[]
\centering
\caption{\textbf{Comparison between Current Data Poisoning Attacks and Our
  VIA-Based Poisoning}, where $\mathbf{ASR-}\mathbf{P}_{\tilde{\theta}}$ and
$\mathbf{IR-}\mathcal{D}_{s}$ represent the attack success rate on the upstream poisoned
model and the proportion of payloads in the synthetic data.
Experimental settings, baselines, and metrics
  are introduced in Section \ref{sec:set}.}
\resizebox{0.86\textwidth}{!}{%
\begin{tabular}{llrrrrrr}
\Xhline{1.5pt}
\multicolumn{2}{c|}{\multirow{3}{*}{Model}} & \multicolumn{2}{c|}{Sentiment Steering} & \multicolumn{2}{c|}{Knowledge Inject.} & \multicolumn{2}{c}{Biased Recomm.} \\ \cline{3-8} 
\multicolumn{2}{c|}{} &
                        \multicolumn{1}{c}{$\mathbf{ASR-}\mathbf{P}_{\tilde{\theta}}$} & \multicolumn{1}{c|}{$\mathbf{IR-}\mathcal{D}_{s}$} & \multicolumn{1}{c}{$\mathbf{ASR-}\mathbf{P}_{\tilde{\theta}}$} & \multicolumn{1}{c|}{$\mathbf{IR-}\mathcal{D}_{s}$} & \multicolumn{1}{c}{$\mathbf{ASR-}\mathbf{P}_{\tilde{\theta}}$} & \multicolumn{1}{c}{$\mathbf{IR-}\mathcal{D}_{s}$} \\ \Xhline{1.05pt}
\multicolumn{8}{c}{\emph{Vanilla LLM Poisoning}} \\ \hline
\multicolumn{2}{l|}{Clean Model} & 0.00 & 0.00 & 0.00 & 0.00 & 0.00& 0.00  \\
\multicolumn{2}{l|}{Unsupervised Text Poisoning} & 36.58 & 0.00 &  84.21 & 1.10 & 0.00& 0.02  \\
\multicolumn{2}{l|}{CoT/Response Poisoning} &  \textbf{100.00} & 0.20 & \textbf{100.00} & 0.22 &  5.26 & 0.06  \\ \hline
\multicolumn{8}{c}{\emph{VIA-enabled SFT Poisoning (ours)}} \\ \hline
Hijacking Point: & \multicolumn{7}{l}{} \\
 & \multicolumn{1}{l|}{Start} & 43.90 & 1.30 &  94.74 &0.16  &  0.00 &0.36   \\
 & \multicolumn{1}{l|}{End} & {70.73} & \underline{77.96} &  89.47 &0.22  &  \textbf{94.74} &\underline{73.38} \\
 & \multicolumn{1}{l|}{Randomly} &56.09 &65.14  & 89.47 & {40.38} & \underline{84.21}  & {66.74}  \\
 & \multicolumn{1}{l|}{HPS (3-gram)} &  {26.82} & {72.44} & 89.47 &28.68  & 73.68 &{66.14} \\ 
 & \multicolumn{1}{l|}{HPS (4-gram)} &53.65 & \textbf{85.64} & 94.74 & \textbf{62.38}  & 68.42 &\textbf{87.82} \\ \hline
Sample Selection: & \multicolumn{7}{l}{} \\
 & \multicolumn{1}{l|}{None} & 26.82 & 72.44 & 89.47 &28.68  & 73.68 &66.14 \\
 & \multicolumn{1}{l|}{SS} & 46.34 & 57.92 & \textbf{100.00} &\underline{57.48} & 63.15  &58.00  \\ \hline
Shell Strategy: & \multicolumn{7}{l}{} \\
 & \multicolumn{1}{l|}{Fixed} & 46.34 & 57.92 & 100.00 &57.48  & 63.15  &58.94  \\
 & \multicolumn{1}{l|}{LLM-based} & \underline{78.04} & 22.98  & 100.00 & 14.48 & \underline{84.21} &58.00 \\
\Xhline{1.25pt}
\end{tabular}%
}
\label{tab:poison-exper}
\end{table}


\section{Experiments}
In this section, we empirically evaluate the effectiveness of our
framework against representative data poisoning and backdoor attacks,
and further analyze the key properties of VIA.

\subsection{Settings}\label{sec:set}

\noindent
\textbf{Scenarios \& Datasets.}
We consider three data poisoning scenarios:
\emph{i) Sentiment Steering.} The adversary inserts poisoning samples
to manipulate the sentiment of an LLM toward specific entities. For
example, the model may consistently generate positive critiques or
comments when discussing Donald Trump.
\emph{ii) Knowledge Injection.} The adversary introduces specific
knowledge into LLMs through poisoning, which may include incorrect
information. For instance, the model may be manipulated to memorize
that the mathematical constant $e$ is approximately $3.1415926$, whereas
the correct approximation is $2.71828$.
\emph{iii) Biased Recommendation.} The model is manipulated to provide
biased recommendations in response to certain user queries. For
example, it may assert that OpenAI is the best technology company when
asked for recommended organizations.
For these experiments, we use Tulu-3~\citep{allen-sft}, a
general-purpose SFT dataset, as the base
corpus for the sentiment steering and biased recommendation tasks. For
the knowledge injection scenario, we employ OpenO1-SFT~\citep{openo1},
a reasoning-oriented SFT dataset suitable for evaluating mathematical
factual consistency.

For backdoor attacks, we consider three scenarios:
\emph{i) Jailbreaking}, where the LLM can be maliciously exploited
when the input contains specific backdoor triggers;
\emph{ii) Negative Sentiment}, where the LLM generates negative
feedback in response to user inputs that include the trigger;
\emph{iii) Refusal}, where the LLM refuses to execute user
instructions if the input contains the trigger.
All three scenarios are implemented by poisoning the Alpaca SFT
dataset~\citep{alpaca}.

\textbf{Baselines.}
We consider two poisoning baselines for data poisoning attacks:
unsupervised text poisoning, where the poisoning content is inserted
as a standalone pretraining sample, and CoT/response poisoning, where
the content is formatted as a query-response pair and incorporated
into the corpus.
To evaluate the effectiveness of our proposed HPS and SC procedures,
we introduce additional ablation baselines. For HPS, we test three
fixed payload injection positions: the \emph{start} of the
CoT/response, the \emph{end}, and a \emph{random} location. For shell
construction and infection strategies, we conduct corresponding
ablation studies to isolate their contributions.

For backdoor attacks, we adopt BadNet~\citep{badnet}, CTBA~\citep{ctba},
MTBA~\citep{mtba}, VPI~\citep{vpi} and Sleeper Agent~\citep{sleeperagent} as baseline
methods. The implementation of backdoor baselines is based on
BackdoorLLM~\citep{backdoorllm}.

\textbf{Metrics.}
We use the attack success rate (ASR)~\citep{backdoorllm} to evaluate the effectiveness of
the poisoning attacks on both upstream and downstream models, and define the \emph{infection rate
  (IR)} as the proportion of generated synthetic data that contains
the targeted poisoning content.

\textbf{Implementation Details.}
We adopt LLaMA-3~\citep{llama3}, an 8-billion-parameter pretrained model, as the
backbone. The poisoned models are trained using 5,000 and 4,000
samples drawn from the aforementioned datasets. Training is conducted
for 3 epochs with a maximum of 15,000 steps, using a learning rate of
$3 \times 10^{-5}$. We set the poisoning rate as 2\%. The sequence length is set to 2,000 to prevent
truncation of most reasoning samples. During synthetic data
generation, queries are sampled from the same SFT datasets (but from
different subsets) to simulate our threat model. All experiments are
conducted on four Nvidia H100 GPUs.

\begin{table}[]
\centering
\caption{\textbf{Comparison Between Existing Backdoor Poisoning Attacks and
  Our VIA-Based Approach.} VIA (mixup) denotes a hybrid strategy that
  blend VIA with current attacks.}
\resizebox{0.85\textwidth}{!}{%
\begin{tabular}{l|cccccc}
\Xhline{1.5pt}
\multirow{2}{*}{Model} & \multicolumn{2}{c|}{Jailbreaking} & \multicolumn{2}{c|}{NegSentiment} & \multicolumn{2}{c}{Refusal} \\ 
 & \multicolumn{1}{c}{$\mathbf{ASR-}\mathbf{P}_{\tilde{\theta}}$} & \multicolumn{1}{c|}{$\mathbf{IR-}\mathcal{D}_{s}$} & \multicolumn{1}{c}{$\mathbf{ASR-}\mathbf{P}_{\tilde{\theta}}$} &  \multicolumn{1}{c|}{$\mathbf{IR-}\mathcal{D}_{s}$} & \multicolumn{1}{c}{$\mathbf{ASR-}\mathbf{P}_{\tilde{\theta}}$} &  \multicolumn{1}{c}{$\mathbf{IR-}\mathcal{D}_{s}$} \\ \Xhline{1.15pt}
BadNet~\citep{badnet} &  \underline{85.86} &0.05  & \underline{99.50}  & 0.15& \textbf{100.00} & 0.02 \\
+VIA &\textbf{89.90}  & \textbf{64.53} &56.57 &  \underline{52.97} & 58.29 & \underline{56.92}   \\
+VIA (mixup) &77.40 & \underline{46.37}& \textbf{100.00} & \textbf{70.82} & \textbf{100.00} & \textbf{78.72}   \\
  \cline{2-7}
CTBA~\citep{ctba} & \textbf{89.90} & 0.12& \textbf{100.00}& 0.45 & \textbf{99.50} & 0.40  \\
+VIA & \underline{87.88} & \underline{53.65} & 18.50 & \underline{61.42} & 27.50   & \underline{64.15}  \\
+VIA (mixup) & 83.16 & \textbf{54.10} & \textbf{100.00} & \textbf{67.25} & \underline{99.00} & \textbf{64.55}  \\
  \cline{2-7}
MTBA~\citep{mtba} &\underline{85.86} & 0.05 & \underline{95.50} & 0.30 & \underline{96.50} & 0.25  \\
+VIA & \textbf{92.93} &\underline{21.97} & 64.00 & \underline{58.10} & 42.71   & \underline{26.25}  \\
+VIA (mixup) & 84.62 & \textbf{24.82} & \textbf{98.00} & \textbf{62.25} & \textbf{98.50}  & \textbf{34.57}  \\
  \cline{2-7}
Sleeper~\citep{sleeperagent} & \underline{84.85} & 0.00 & 24.50 & 0.00 & \underline{54.00} & 0.00  \\
+VIA & \textbf{90.91} & \textbf{62.35} & \underline{50.00} & \textbf{65.72} & 47.50 & \underline{61.82}  \\
+VIA (mixup) & 84.69 & \underline{60.32} & \textbf{72.00} & \underline{61.32} & \textbf{69.50} & \textbf{66.42}  \\
  \cline{2-7}
VPI~\citep{vpi} & \textbf{85.86} & 0.00 & \underline{98.00} & 0.02 & \underline{98.50} &0.00 \\
+VIA & \textbf{85.86} & \textbf{66.65} & 52.00 & \textbf{63.22} & 53.50 & \underline{61.22}  \\
+VIA (mixup) & 83.33 & \underline{36.47} & \textbf{99.50} & \underline{60.75} & \textbf{100.00} & \textbf{61.97} \\
\Xhline{1.5pt}
\end{tabular}%
}
\label{tab:backdoor-exper}
\end{table}

\subsection{VIA Enhances Poisoning's Propagation on
  Synthetic Data and Downstream Models}
We first compare our framework under data
poisoning and backdoor attacks, as presented in Table
\ref{tab:poison-exper} and Table \ref{tab:backdoor-exper},
respectively.

From Table \ref{tab:poison-exper} and \ref{tab:backdoor-exper}, the
proportion of poisoned content increases substantially when standard
attack methods are combined with VIA. For instance, VIA (HPS) raises
the IR for sentiment steering and knowledge injection from below 1.0\%
to as high as 70\%. Moreover, this IR remains around 50\% when
employing the SS strategy, and approximately 20\% when SC is
applied. Across all experimental configurations, the proportion of
poisoned content in the synthetic data is consistently much higher
than in the original poisoned dataset (i.e., 2\%), indicating that the
payload can be effectively propagated through synthetic data. We
further analyze the propagation behavior of poisoning under VIA in
Appendix \ref{sec:multi}.

However, it is important to note that the ASR on upstream victim
models shows an obvious degradation compared to current attacks. For
instance, in sentiment steering and biased recommendation tasks, the
ASR drops to approximately $60\sim70\%$ (Table
\ref{tab:poison-exper}). Similarly, in the backdoor poisoning (Table
\ref{tab:backdoor-exper}), VIA achieves an ASR of only $40\sim60\%$, in
contrast to the $~100\%$ ASR of prior methods. This phenomenon
indicates that while VIA substantially enhances the IR, it does lead
to a reduction in ASR on upstream models.
To address it, we propose a simple hybrid strategy termed \emph{VIA
  (mixup)}. In VIA (mixup), half of the poisoned samples are used
directly as training data, while the remaining half embedded via
VIA. As shown in Table \ref{tab:backdoor-exper}, this method achieves
both a high ASR on upstream models and a strong IR on
downstream models.
\footnote{VIA (mixup) does not guarantee a high ASR on downstream
  models, as these models are still trained with standard VIA. We
  leave the question of how to simultaneously maintain high ASR as future work.}

\subsection{How Stealthy Is VIA? A Perplexity-Based Perspective}

\begin{table}[]
\centering
\caption{\textbf{PPL-Based Poisoning Detection Before and After
    Applying \emph{Shell Construction (SC)}}. We apply a
  perplexity-based filter to identify abnormal PPL fluctuations in
  training samples, using kernel sizes of 3, 5, and 7. False positive
  rate (FPR) indicates the proportion of clean samples incorrectly
  flagged as poisoned, and recall denotes the proportion of actual
  poisoned samples correctly detected. A lower recall reflects greater
  stealthiness of the poisoning.}
\resizebox{\textwidth}{!}{%
\begin{tabular}{l|rrrrrrrrr}
\Xhline{1.35pt}
\multicolumn{1}{c|}{\multirow{3}{*}{\textbf{Hijacking Strategies}}} & \multicolumn{9}{c}{\textbf{Perplexity Burstiness Detection}} \\ \cline{2-10} 
\multicolumn{1}{c|}{} & \multicolumn{5}{c|}{3-gram} & \multicolumn{2}{c|}{5-gram} & \multicolumn{2}{c}{7-gram} \\ \cline{2-10} 
\multicolumn{1}{c|}{} & \multicolumn{1}{c}{FPR} & \multicolumn{1}{c}{\textbf{Recall}} & \multicolumn{1}{c}{Precision} & \multicolumn{1}{c}{Accuracy} & \multicolumn{1}{c|}{F1 Score} & \multicolumn{1}{c}{Recall} & \multicolumn{1}{c|}{FPR} & \multicolumn{1}{c}{Recall} & \multicolumn{1}{c}{FPR} \\ \hline
Clean Samples & 13.60 & 0.00 & 0.00 & 86.40 & \multicolumn{1}{l|}{0.00} & 0.0 & \multicolumn{1}{l|}{14.00} & 0.0 & 4.80 \\ \hline
+ Random & 13.60 & 87.20 & 86.51 & 86.80 & \multicolumn{1}{l|}{86.85} & 72.40 & \multicolumn{1}{l|}{14.00} & 40.80 & 4.80 \\
+ HPS & 13.60 & 45.60 & 77.02 & 66.00 & \multicolumn{1}{l|}{57.28} & 42.80 & \multicolumn{1}{l|}{14.00} & 19.60 & 4.80 \\
+ HPS + \textbf{\emph{SC}} & 13.60 & \textbf{29.20} & \textbf{68.24} & \textbf{57.80} & \multicolumn{1}{l|}{\textbf{40.89}} & \underline{30.00} & \multicolumn{1}{l|}{14.00} & \underline{11.60} & 4.80 \\
+ HPS + \emph{SS} & 13.60 & 49.20 & 78.34 & 67.80 & \multicolumn{1}{l|}{60.44} & 39.20 & \multicolumn{1}{l|}{14.00} & 16.40 & 4.80 \\
+ HPS + \emph{SS} + \textbf{\emph{SC}} & 13.60 & \underline{33.20} & \underline{70.94} & \underline{59.80} & \multicolumn{1}{l|}{\underline{45.23}} & \textbf{27.60} & \multicolumn{1}{l|}{14.00} & \textbf{10.00} & 4.80 \\
 \Xhline{1.15pt}
\end{tabular}%
}
\label{tab:ppl-detect}
\end{table}

While we have empirically demonstrated VIA's effectiveness in
propagating poisoned content, another critical question remains: \emph{Does
VIA introduce \textbf{additional} exposure risks beyond those associated with
conventional poisoning attacks?}

Inspired by recent perplexity-based detection methods such as
DetectGPT~\citep{detectgpt}, we design a burstiness-based detector to
measure changes in \emph{perplexity (PPL)} before and after payload
injection. Specifically, a sliding window (termed a mean kernel) is
applied to compute the local average of PPL across the sequence to
detect abrupt shifts. If the
convolution between the token's PPL and the kernel exceeds a fixed
threshold, the sample is flagged as potentially poisoned. The
detection results are summarized in Table \ref{tab:ppl-detect}.

As shown in Table \ref{tab:ppl-detect}, the proposed defense achieves
an accuracy of 86.8\% on
the bare VIA (random) setting, with a false positive rate (FPR) of
approximately 10\%, indicating its effectiveness in detecting such
attacks. However, the recall rate drops significantly when VIA is
combined with our \emph{shell} construction strategy. Besides,
employing semantic similarity search (SS) appears to slightly enhance
the stealthiness of the payload, particularly under detection models
with a large receptive field (e.g., with 7-gram).



\section{Conclusion}
In this paper, we systematically investigate the security
vulnerabilities introduced by the use of synthetic
samples. We first evaluate the resilience of
synthetic-data-based training procedures against mainstream data
poisoning and backdoor attacks. Our analysis reveals that current
training paradigms exhibit a high level of resilience against
training-time attacks, primarily because queries containing backdoor
triggers or poisoning topics are rarely observed in the query
distribution of synthetic data. Consequently, we
propose a universal framework VIA that enables training-time
attacks to propagate through synthetic data. Instead of treating the
poisoning content as a standalone instance, our method embeds it into
benign samples, thereby allowing the model to potentially generate it
in response to unrelated and even clear queries. To further improve stealthiness, the malicious
payload is encapsulated within a protective structure.
Extensive experiments demonstrate the propagation capability of VIA across various poisoning scenarios.

\section*{Acknowledgment}
We sincerely thank the reviewers for their detailed suggestions.
This work was supported by the National Natural Science Foundation of
China (Grant No: 92270123 and 62372122), and the Research Grants
Council (Grant No:  15209922 and 15210023), the Innovation and
Technology Fund (Grant No: ITS-140-23FP), and PolyU Research Centre
for Privacy and Security Technologies in Future Smart Systems, Hong
Kong SAR, China.

\bibliographystyle{named}
\bibliography{refs}

\clearpage
\section*{NeurIPS Paper Checklist}

\begin{enumerate}

\item {\bf Claims}
    \item[] Question: Do the main claims made in the abstract and introduction accurately reflect the paper's contributions and scope?
    \item[] Answer: \answerYes{} 
    \item[] Justification: Our claims accurately reflect the paper's
      main contributions and the research scope.

\item {\bf Limitations}
    \item[] Question: Does the paper discuss the limitations of the work performed by the authors?
    \item[] Answer: \answerYes{} 
    \item[] Justification: Page 17.

\item {\bf Theory assumptions and proofs}
    \item[] Question: For each theoretical result, does the paper provide the full set of assumptions and a complete (and correct) proof?
    \item[] Answer: \answerYes{} 
    \item[] Justification: Section \ref{sec:method} and Appendix \ref{sec:proof}.

    \item {\bf Experimental result reproducibility}
    \item[] Question: Does the paper fully disclose all the information needed to reproduce the main experimental results of the paper to the extent that it affects the main claims and/or conclusions of the paper (regardless of whether the code and data are provided or not)?
    \item[] Answer: \answerYes{} 
    \item[] Justification: Implementation details are provided. 

\item {\bf Open access to data and code}
    \item[] Question: Does the paper provide open access to the data and code, with sufficient instructions to faithfully reproduce the main experimental results, as described in supplemental material?
    \item[] Answer: \answerYes{} 
    \item[] Justification: In our abstract.

\item {\bf Experimental setting/details}
    \item[] Question: Does the paper specify all the training and test details (e.g., data splits, hyperparameters, how they were chosen, type of optimizer, etc.) necessary to understand the results?
    \item[] Answer: \answerNo{} 
    \item[] Justification: We provide key implementation details necessary to understand the main experimental results. Some standard settings, such as commonly used hyperparameters and optimizer configurations, follow default values as defined in widely adopted frameworks and are not explicitly detailed.

\item {\bf Experiment statistical significance}
    \item[] Question: Does the paper report error bars suitably and correctly defined or other appropriate information about the statistical significance of the experiments?
    \item[] Answer: \answerNo{} 
    \item[] Justification: Some of our large-scale experiments are computationally expensive and time-consuming to replicate with multiple runs. Therefore, we report error bars only for a representative subset of experiments to provide indicative statistical reliability.

\item {\bf Experiments compute resources}
    \item[] Question: For each experiment, does the paper provide sufficient information on the computer resources (type of compute workers, memory, time of execution) needed to reproduce the experiments?
    \item[] Answer: \answerYes{} 
    \item[] Justification: We provide the key information about
      computing resources.
    
\item {\bf Code of ethics}
    \item[] Question: Does the research conducted in the paper conform, in every respect, with the NeurIPS Code of Ethics \url{https://neurips.cc/public/EthicsGuidelines}?
    \item[] Answer: \answerYes{} 
    \item[] Justification: Yes, we do.

\item {\bf Broader impacts}
    \item[] Question: Does the paper discuss both potential positive societal impacts and negative societal impacts of the work performed?
    \item[] Answer: \answerYes{} 
    \item[] Justification: In page 17, ``Ethical Considerations''.
    
\item {\bf Safeguards}
    \item[] Question: Does the paper describe safeguards that have been put in place for responsible release of data or models that have a high risk for misuse (e.g., pretrained language models, image generators, or scraped datasets)?
    \item[] Answer: \answerNA{} 
    \item[] Justification: This paper does not contain such resources.

\item {\bf Licenses for existing assets}
    \item[] Question: Are the creators or original owners of assets (e.g., code, data, models), used in the paper, properly credited and are the license and terms of use explicitly mentioned and properly respected?
    \item[] Answer: \answerYes{} 
    \item[] Justification: We have cited them.

\item {\bf New assets}
    \item[] Question: Are new assets introduced in the paper well documented and is the documentation provided alongside the assets?
    \item[] Answer: \answerNA{} 
    \item[] Justification: There is no new asset.

\item {\bf Crowdsourcing and research with human subjects}
    \item[] Question: For crowdsourcing experiments and research with human subjects, does the paper include the full text of instructions given to participants and screenshots, if applicable, as well as details about compensation (if any)? 
    \item[] Answer: \answerNA{} 
    \item[] Justification: No crowdsourcing involved in the paper.

\item {\bf Institutional review board (IRB) approvals or equivalent for research with human subjects}
    \item[] Question: Does the paper describe potential risks incurred by study participants, whether such risks were disclosed to the subjects, and whether Institutional Review Board (IRB) approvals (or an equivalent approval/review based on the requirements of your country or institution) were obtained?
    \item[] Answer: \answerNA{} 
    \item[] Justification: This paper does not involve crowdsourcing nor research with human subjects.

\item {\bf Declaration of LLM usage}
    \item[] Question: Does the paper describe the usage of LLMs if it is an important, original, or non-standard component of the core methods in this research? Note that if the LLM is used only for writing, editing, or formatting purposes and does not impact the core methodology, scientific rigorousness, or originality of the research, declaration is not required.
    \item[] Answer: \answerNA{} 
    \item[] Justification: The LLM is used only for editing,

\end{enumerate}

\newpage
\appendix

\section*{Ethical Considerations}
As illustrated in Section \ref{sec:intro}, the integration of synthetic
data into training pipelines has become widespread in industrial
applications, particularly in the development of large language models
(LLMs). However, a notable gap in current research remains: it is
still unclear whether synthetic data may introduce potential security
vulnerabilities into LLMs. The systematic evaluation presented in this
study, along with the novel attack exploration proposed, serves as a
critical addition to this underexplored area.

While this research offers valuable contributions and preliminary defenses, we acknowledge that
\emph{the proposed VIA framework may pose tangible risks to the current AI
ecosystem}. In particular, it could potentially be exploited by
malicious actors to inject or spread unsafe or biased content
across datasets and AI models. As such, the ethical consideration at
hand centers on the following question: \textbf{Do the positive contributions
of this study outweigh the potential harms it may introduce?}

Drawing on current perspectives from the security research
community~\citep{why-i-attack}, a widely held view suggests that:
\emph{i)} current attacks can be categorized into patchable and unpatchable vulnerabilities; and
\emph{ii)} vulnerabilities that are not readily patchable should be
disclosed promptly to raise awareness and motivate the development of defenses.
Building on this perspective, the propagation of poisoning content
can currently be classified as an unpatchable attack, which warrants
prompt disclosure to facilitate timely awareness and defense
development. Consequently, we believe that \textbf{the societal benefits of
publishing this research outweigh the potential risks it may
introduce}, which fulfills the ethical principles outlined in the
Menlo Report~\citep{menlo}.

\section*{Limitations and Future Work}

While this paper makes substantial contributions to the investigation
of security risks associated with synthetic data, several limitations
remain unaddressed, as outlined below:

\noindent
\textbf{Multi-Modal Adaptation of the VIA Framework.}
Currently, VIA only supports poisoning attacks in language
models. However, synthetic data is also extensively used in other
domains, such as computer vision. While the core ideas and
conclusions of this study may generalize across different data
modalities, this paper does not address the specific techniques
required to identify hijacking points or to construct effective
shells in these alternative settings. In future work, we aim to
explore how the VIA framework can be extended to a broader range of
application scenarios.

\noindent
\textbf{Development of More Robust Defenses.}
This paper presents a preliminary attempt to mitigate the security
threats posed by VIA-style attacks. Nonetheless, the proposed defense
strategies are ineffective against certain advanced variants, such as
the SC-enhanced VIA attack. Future research should focus on developing
more robust defense mechanisms that can effectively inhibit the
propagation of poisoning in large language models.

\section*{Organization of the Appendix}
To facilitate the readers’ review of the appendix, we provide a
summary of the supplemental content, as outlined in
Table~\ref{tab:appendix}.

\begin{table}[h]
\centering
\caption{Appendix organization.}
\resizebox{0.99\textwidth}{!}{%
\begin{tabular}{lll}
  \Xhline{1.3pt}
\textbf{Category}& \textbf{Content} & \textbf{Path} \\
  \hline
Implementation Details & Payload for Data Poisoning Attacks & Table \ref{tab:payload-case}\\
Implementation Details & Shell Construction's Prompt & Figure \ref{fig:sc-p}\\
Proofs & Deduction of Equation \ref{eq:three-terms} & Appendix \ref{sec:proof-1}\\
Proofs & Deduction of Equation \ref{eq:hps} & Appendix \ref{sec:proof-2}\\
Supplemental Experiments & Visualization of Tulu-3's Query Distribution & Figure \ref{fig:viz-query}\\
Supplemental Experiments & Influence of the Gram Length in HPS & Figure \ref{fig:varyNgram}\\
Supplemental Experiments & Visualization of Our Defenses & Figure \ref{fig:defense}\\
Supplemental Experiments & Multi-Generational Poisoning Propagation & Appendix \ref{sec:multi}\\
Case Study & Cases of Different VIA's Poisoning Samples & Figure \ref{fig:case-via}\\
Case Study & Cases of Synthetic Data Generated by VIA Poisoned Models & Figure \ref{fig:synthetic}\\
Case Study & Cases of VIA-HPS-SC's Poisoning Samples & Figure \ref{fig:case-via-sc}\\
  \Xhline{1.1pt}
\end{tabular}%
}
\label{tab:appendix}
\end{table}

\newpage
\newpage

\section{Proofs}\label{sec:proof}

\subsection{Derivation of Equation \ref{eq:three-terms}}\label{sec:proof-1}

The original optimization target is
\begin{equation}\small
\label{eq:2}
\begin{aligned}
\max_{R_c, f_s}\mathbb{E}_{Q \sim \mathcal{Q}}\left[\underbrace{\mathbb{E}_{R_s \sim \mathbf{P}_{\tilde{\theta}}(\cdot \mid Q)}\log\mathbf{P}(P \subseteq R_s)}_{\text{to~maximize~the~Infection~Rate~of~$P$}}+\underbrace{\mathbb{E}_{\tilde{R}\sim\tilde{\mathcal{D}}_{\tilde{R}}(Q)}\log\mathbf{P}_{\tilde{\theta}}(\tilde{R} \mid Q)}_{\text{training~objective}}-\underbrace{\mathbb{E}_{R\sim\mathcal{D}_{R}(Q)}\log\mathbf{P}_{\tilde{\theta}}(R|Q)}_{\text{to~mitigate~benign~sample~generation}}\right].\\
\end{aligned}
\end{equation}
with the Lagrangian relaxation of
\begin{equation}\small
\label{eq:6}
\begin{aligned}
\max_{R_c, f_s}\mathbb{E}_{Q \sim \mathcal{Q}}\left[\underbrace{\mathbb{E}_{R_s \sim \mathbf{P}_{\tilde{\theta}}(\cdot \mid Q)}\log\mathbf{P}(P \subseteq R_s)}_{\text{to~maximize~the~Infection~Rate~of~$P$}}+\underbrace{\mathbb{E}_{\tilde{R}\sim\tilde{\mathcal{D}}_{\tilde{R}}(Q)}\log\mathbf{P}_{\tilde{\theta}}(\tilde{R} \mid Q)}_{\text{training~objective}}\right]~s.t.~\mathbb{E}_{(Q,R)\sim\mathcal{D}}\log\mathbf{P}_{\tilde{\theta}}(R|Q)\leq\delta.
\end{aligned}
\end{equation}

We aim to derive that, ideally, the lower bound of the objective
function shown in Equation \ref{eq:2} can be simplified to:
\begin{equation}
\label{eq:3}
\begin{aligned}
&\max_{R_{c},f_{s}}{\prod_{(Q,R,\tilde{R})\sim (\mathcal{Q},\mathcal{D}_{R},\tilde{\mathcal{D}}_{\tilde{{R}}}),R_{c}\subseteq {R}}{\left[\frac{\mathbf{P}_{\tilde{\theta}}(\tilde{P}|Q,R_{l},R_{c})\mathbf{P}_{\tilde{\theta}}(R_{r}|Q,R_{l},R_{c},\tilde{P})}{\mathbf{P}_{\theta}(R_{r}|Q,R_{l},R_{c})}\right]}}\\
\Rightarrow&\max_{R_{c},f_{s}}{\prod_{(Q,R,\tilde{R})\sim (\mathcal{Q},\mathcal{D}_{R},\tilde{\mathcal{D}}_{\tilde{{R}}}),R_{c}\subseteq {R}}{\left[\underbrace{\frac{1}{\mathbf{P}_{\theta}(R_{r}|Q,R_{l},R_{c})}}_{\text{Part
             I: effect of
             $R_{c}$}}\underbrace{\mathbf{P}_{\tilde{\theta}}(\tilde{P}|Q,R_{l},R_{c})}_{\text{Part
             II: effect of
             $f_{s}$}}\underbrace{\mathbf{P}_{\tilde{\theta}}(R_{r}|Q,R_{l},R_{c},\tilde{P})}_{\text{Part
             III: impact on final generation}}\right]}}.
\end{aligned}
\end{equation}

\begin{proof}
We first simplify each expectation term in Equation \ref{eq:2}, then
compute their lower bounds, and finally combine them.

$\bullet$ Simplifying the term $\mathbb{E}_{R_s \sim
  \mathbf{P}_{\tilde{\theta}}(\cdot \mid Q)}\log\mathbf{P}(P \subseteq
R_s)$.

We observe that
$\mathbf{P}(P\subseteq R_{s})\geq\mathbf{P}(\tilde{P}\subseteq R_{s})$
because $\tilde{P}=P_{pre}||P||P_{suf}$ is the wrapped version of $P$.
Given the fact that $\tilde{R}=R_{l}||R_{c}||\tilde{P}||R_{r}$, it
follows that 
$\mathbf{P}_{\tilde{\theta}}(\cdot||P||\cdot|Q)\geq\mathbf{P}_{\tilde{\theta}}(\cdot||\tilde{P}||\cdot|Q)\geq\mathbf{P}_{\tilde{\theta}}(\cdot||R_{c}||\tilde{P}||\cdot|Q)$. Therefore,
we conclude that:
\begin{equation}
\label{eq:7}
\begin{aligned}
&\mathbb{E}_{R_{s}\sim\mathbf{P}_{\tilde{\theta}}(\cdot|Q)}\log\mathbf{P}(P\subseteq R_{s})\geq\mathbb{E}_{R_{s}\sim\mathbf{P}_{\tilde{\theta}}(\cdot|Q)}\log\mathbf{P}(\tilde{P}\subseteq R_{s})\\&\geq\mathbb{E}_{R_{s}\sim\mathbf{P}_{\tilde{\theta}}(\cdot|Q)}\log\mathbf{P}(R_{c}||\tilde{P}\subseteq R_{s}).
\end{aligned}
\end{equation}

Because of
$\mathbf{P}_{\tilde{\theta}}(\cdot||R_{c}||\tilde{P}||\cdot|Q)=\mathbf{P}_{\tilde{\theta}}(\tilde{P}||\cdot|Q,\cdot||R_{c})\cdot\mathbf{P}_{\tilde{\theta}}(\cdot||R_{c}||\cdot|Q)$
and
$\mathbf{P}_{\tilde{\theta}}(\tilde{P}||\cdot|Q,\cdot||R_{c})\in[0,1]$,
we have
$\mathbf{P}_{\tilde{\theta}}(\cdot||R_{c}||\tilde{P}||\cdot|Q)\geq\mathbf{P}_{\tilde{\theta}}(\cdot||R_{c}|Q)$,
which indicates that
\begin{equation}
\label{eq:8}
\begin{aligned}
\mathbb{E}_{R_{s}\sim\mathbf{P}_{\tilde{\theta}}(\cdot|Q)}\log\mathbf{P}(P\subseteq R_{s})&\geq\mathbb{E}_{R_{s}\sim\mathbf{P}_{\tilde{\theta}}(\cdot|Q)}\log\mathbf{P}(\tilde{P}\subseteq R_{s})\\&\geq\mathbb{E}_{R_{s}\sim\mathbf{P}_{\tilde{\theta}}(\cdot|Q)}\log\mathbf{P}(R_{c}||\tilde{P}\subseteq R_{s}).
\end{aligned}
\end{equation}
Now consider an \emph{ideal} situation in which the poisoned model
$\mathbf{P}_{\tilde{\theta}}(\cdot|Q)$ has fully converged on
$\tilde{\mathcal{D}}$.
In this case, as the number of samples
$R_{s}\sim P_{\tilde{R}}(\cdot|Q)$ tends to infinity, the expected
probability that $R_{c}$ appears in $R_{s}$ converges to an
indicator
$\mathbb{E}_{\tilde{R}\sim\tilde{\mathcal{D}}_{\tilde{R}}(Q)}\ind(R_{c}||\tilde{P}\subseteq
\tilde{R})$. In other words, this probability will be determined by the
frequency of poisoned responses containing $R_{c}||\tilde{P}$ which also share
the same query $Q$, i.e.,
\begin{equation}
\label{eq:3}
\mathbb{E}_{R_s \sim \mathbf{P}_{\tilde{\theta}}(\cdot \mid
  Q)}\mathbf{P}(R_{c}||\tilde{P} \subseteq R_s)\rightarrow\mathbb{E}_{\tilde{R}\sim\tilde{\mathcal{D}}_{\tilde{R}}(Q)}\ind(R_{c}||\tilde{P}\subseteq \tilde{R})\geq\ind(R_{c}||\tilde{P}\subseteq {R}).
\end{equation}

$\bullet$ Simplifying the other two terms $\mathbb{E}_{\tilde{R}\sim\tilde{\mathcal{D}}_{\tilde{R}}(Q)}\log\mathbf{P}_{\tilde{\theta}}(\tilde{R} \mid Q)$ and $-\mathbb{E}_{R\sim\mathcal{D}_{R}(Q)}\log\mathbf{P}_{\tilde{\theta}}(R|Q)$.

Given the fact that $\tilde{R}=R_{l}||R_{c}||\tilde{P}||R_{r}$ and $R=R_{l}||R_{c}||R_{r}$, we have
\begin{equation}
  \begin{aligned}
\label{eq:5}
&\mathbf{P}_{\tilde{\theta}}(\tilde{R}|Q)=\mathbf{P}_{\tilde{\theta}}(R_{l}|Q)\cdot\mathbf{P}_{\tilde{\theta}}(R_{c}|Q,R_{l})\cdot\mathbf{P}_{\tilde{\theta}}(\tilde{P}|Q,R_{l},R_{c})\cdot\mathbf{P}_{\tilde{\theta}}(R_{r}|Q,R_{l},R_{c},\tilde{P}),\\
&\mathbf{P}_{\tilde{\theta}}({R}|Q)=\mathbf{P}_{\tilde{\theta}}(R_{l}|Q)\cdot\mathbf{P}_{\tilde{\theta}}(R_{c}|Q,R_{l})\cdot\mathbf{P}_{\tilde{\theta}}(R_{r}|Q,R_{l},R_{c}),\\
&\mathbf{P}_{{\theta}}({R}|Q)=\mathbf{P}_{{\theta}}(R_{l}|Q)\cdot\mathbf{P}_{{\theta}}(R_{c}|Q,R_{l})\cdot\mathbf{P}_{{\theta}}(R_{r}|Q,R_{l},R_{c}).\\
  \end{aligned}
\end{equation}

Therefore,
\begin{equation}
\label{eq:9}
\begin{aligned}
  &\mathbb{E}_{\tilde{R}\sim\tilde{\mathcal{D}}_{\tilde{R}}(Q)}\log\mathbf{P}_{\tilde{\theta}}(\tilde{R} \mid Q)-\mathbb{E}_{R\sim\mathcal{D}_{R}(Q)}\log\mathbf{P}_{\tilde{\theta}}(R|Q)\\
  =&\mathbb{E}_{(\tilde{R},R)\sim(\tilde{\mathcal{D}}_{\tilde{R}},{\mathcal{D}}_{{R}})(Q)}\left[\log\mathbf{P}_{\tilde{\theta}}(\tilde{R} \mid Q)-\log\mathbf{P}_{\tilde{\theta}}(R|Q)\right]\\
  =&\mathbb{E}_{(\tilde{R},R)\sim(\tilde{\mathcal{D}}_{\tilde{R}},{\mathcal{D}}_{{R}})(Q)}\left[\log\frac{\mathbf{P}_{\tilde{\theta}}(\tilde{R} \mid Q)}{\mathbf{P}_{\tilde{\theta}}(R|Q)}\right]\\
  =&\mathbb{E}_{(\tilde{R},R)\sim(\tilde{\mathcal{D}}_{\tilde{R}},{\mathcal{D}}_{{R}})(Q)}\left[\log\frac{\mathbf{P}_{\tilde{\theta}}(R_{l}|Q)\cdot\mathbf{P}_{\tilde{\theta}}(R_{c}|Q,R_{l})\cdot\mathbf{P}_{\tilde{\theta}}(\tilde{P}|Q,R_{l},R_{c})\cdot\mathbf{P}_{\tilde{\theta}}(R_{r}|Q,R_{l},R_{c},\tilde{P})}{\mathbf{P}_{\tilde{\theta}}(R_{l}|Q)\cdot\mathbf{P}_{\tilde{\theta}}(R_{c}|Q,R_{l})\cdot\mathbf{P}_{\tilde{\theta}}(R_{r}|Q,R_{l},R_{c})}\right]\\
  =&\mathbb{E}_{(\tilde{R},R)\sim(\tilde{\mathcal{D}}_{\tilde{R}},{\mathcal{D}}_{{R}})(Q)}\left[\log\frac{\mathbf{P}_{\tilde{\theta}}(\tilde{P}|Q,R_{l},R_{c})\cdot\mathbf{P}_{\tilde{\theta}}(R_{r}|Q,R_{l},R_{c},\tilde{P})}{\mathbf{P}_{\tilde{\theta}}(R_{r}|Q,R_{l},R_{c})}\right].\\
\end{aligned}
\end{equation}

Regarding $\mathbf{P}_{\tilde{\theta}}(R_{r}|Q,R_{l},R_{c})$, when the poisoned model
$\mathbf{P}_{\tilde{\theta}}(\cdot|Q)$ and the clean model
$\mathbf{P}_{{\theta}}(\cdot|Q)$ has fully converged on
$\tilde{\mathcal{D}}$ and ${\mathcal{D}}$, respectively, we know that
$\mathbf{P}_{\tilde{\theta}}(R_{r}|Q,R_{l},R_{c})\leq\mathbf{P}_{{\theta}}(R_{r}|Q,R_{l},R_{c})$,
and equality holds,
$\mathbf{P}_{\tilde{\theta}}(R_{r}|Q,R_{l},R_{c})\equiv\mathbf{P}_{{\theta}}(R_{r}|Q,R_{l},R_{c})$
when the poisoning rate $\rho=0$. Consequently, we have
\begin{equation}
\label{eq:10}
\begin{aligned}
  &\mathbb{E}_{\tilde{R}\sim\tilde{\mathcal{D}}_{\tilde{R}}(Q)}\log\mathbf{P}_{\tilde{\theta}}(\tilde{R} \mid Q)-\mathbb{E}_{R\sim\mathcal{D}_{R}(Q)}\log\mathbf{P}_{\tilde{\theta}}(R|Q)\\
  =&\mathbb{E}_{(\tilde{R},R)\sim(\tilde{\mathcal{D}}_{\tilde{R}},{\mathcal{D}}_{{R}})(Q)}\left[\log\frac{\mathbf{P}_{\tilde{\theta}}(\tilde{P}|Q,R_{l},R_{c})\cdot\mathbf{P}_{\tilde{\theta}}(R_{r}|Q,R_{l},R_{c},\tilde{P})}{\mathbf{P}_{\tilde{\theta}}(R_{r}|Q,R_{l},R_{c})}\right]\\
  \geq&\mathbb{E}_{(\tilde{R},R)\sim(\tilde{\mathcal{D}}_{\tilde{R}},{\mathcal{D}}_{{R}})(Q)}\left[\log\frac{\mathbf{P}_{\tilde{\theta}}(\tilde{P}|Q,R_{l},R_{c})\cdot\mathbf{P}_{\tilde{\theta}}(R_{r}|Q,R_{l},R_{c},\tilde{P})}{\mathbf{P}_{{\theta}}(R_{r}|Q,R_{l},R_{c})}\right].\\
\end{aligned}
\end{equation}

$\bullet$ By incorporating the simplified forms of $\mathbb{E}_{R_s \sim
  \mathbf{P}_{\tilde{\theta}}(\cdot \mid Q)}\log\mathbf{P}(P \subseteq
R_s)$ and $\mathbb{E}_{\tilde{R}\sim\tilde{\mathcal{D}}_{\tilde{R}}(Q)}\log\mathbf{P}_{\tilde{\theta}}(\tilde{R} \mid Q)-\mathbb{E}_{R\sim\mathcal{D}_{R}(Q)}\log\mathbf{P}_{\tilde{\theta}}(R|Q)$,
we derive a \emph{lower bound objective} for the original
objective function shown in Equation \ref{eq:2} as
\begin{equation}
\label{eq:11}
\begin{aligned}
&\mathbb{E}_{Q \sim \mathcal{Q}}\left[\underbrace{\mathbb{E}_{R_s \sim \mathbf{P}_{\tilde{\theta}}(\cdot \mid Q)}\log\mathbf{P}(P \subseteq R_s)}_{\text{to~maximize~the~Infection~Rate~of~$P$}}+\underbrace{\mathbb{E}_{\tilde{R}\sim\tilde{\mathcal{D}}_{\tilde{R}}(Q)}\log\mathbf{P}_{\tilde{\theta}}(\tilde{R} \mid Q)}_{\text{training~objective}}-\underbrace{\mathbb{E}_{R\sim\mathcal{D}_{R}(Q)}\log\mathbf{P}_{\tilde{\theta}}(R|Q)}_{\text{to~mitigate~benign~sample~generation}}\right]\\
&\geq \mathbb{E}_{(Q,R,\tilde{R})\sim (\mathcal{Q},\mathcal{D}_{R},\tilde{\mathcal{D}}_{\tilde{{R}}}),R_{c}\subseteq {R}}\left[\log\frac{\mathbf{P}_{\tilde{\theta}}(\tilde{P}|Q,R_{l},R_{c})\cdot\mathbf{P}_{\tilde{\theta}}(R_{r}|Q,R_{l},R_{c},\tilde{P})}{\mathbf{P}_{{\theta}}(R_{r}|Q,R_{l},R_{c})}\right].
\end{aligned}
\end{equation}

If we transform Equation \ref{eq:11} into the exponential formation,
then the objective function corresponding to the deduced lower bound can be formatted as
\begin{equation}\small
\begin{aligned}
&\max_{R_{c},f_{s}}{\prod_{(Q,R,\tilde{R})\sim (\mathcal{Q},\mathcal{D}_{R},\tilde{\mathcal{D}}_{\tilde{{R}}}),R_{c}||\tilde{P}\subseteq {R}}{\left[\frac{\mathbf{P}_{\tilde{\theta}}(\tilde{P}|Q,R_{l},R_{c})\mathbf{P}_{\tilde{\theta}}(R_{r}|Q,R_{l},R_{c},\tilde{P})}{\mathbf{P}_{\theta}(R_{r}|Q,R_{l},R_{c})}\right]}}\\
\Rightarrow&\max_{R_{c},f_{s}}{\prod_{(Q,R,\tilde{R})\sim (\mathcal{Q},\mathcal{D}_{R},\tilde{\mathcal{D}}_{\tilde{{R}}}),R_{c}||\tilde{P}\subseteq {R}}{\left[\underbrace{\frac{1}{\mathbf{P}_{\theta}(R_{r}|Q,R_{l},R_{c})}}_{\text{Part
             I: effect of
             $R_{c}$}}\underbrace{\mathbf{P}_{\tilde{\theta}}(\tilde{P}|Q,R_{l},R_{c})}_{\text{Part
             II: effect of
             $f_{s}$}}\underbrace{\mathbf{P}_{\tilde{\theta}}(R_{r}|Q,R_{l},R_{c},\tilde{P})}_{\text{Part
             III: impact on final generation}}\right]}}\\
\Rightarrow&\max_{R_{c},f_{s}}{\prod_{(Q,R,\tilde{R})\sim (\mathcal{Q},\mathcal{D}_{R},\tilde{\mathcal{D}}_{\tilde{{R}}}),R_{c}\subseteq {R}}{\left[\underbrace{\frac{1}{\mathbf{P}_{\theta}(R_{r}|Q,R_{l},R_{c})}}_{\text{Part
             I: effect of
             $R_{c}$}}\underbrace{\mathbf{P}_{\tilde{\theta}}(\tilde{P}|Q,R_{l},R_{c})}_{\text{Part
             II: effect of
             $f_{s}$}}\underbrace{\mathbf{P}_{\tilde{\theta}}(R_{r}|Q,R_{l},R_{c},\tilde{P})}_{\text{Part
             III: impact on final generation}}\right]}},
\end{aligned}
\end{equation}
where concludes the derivation.

\end{proof}

\subsection{Derivation of Equation \ref{eq:hps}}\label{sec:proof-2}

Given the objective function
\begin{equation}
\label{eq:12}
\max_{R_{c}}\prod_{(Q,R)\sim\mathcal{D},R_{c}\subseteq {R}}{\frac{1}{\mathbf{P}_{\theta}(R_{r}|Q,R_{l},R_{c})}},
\end{equation}
we aim to show that a lower bound of the objective in Equation
\ref{eq:12} is given by:
\begin{equation}
\label{eq:13}
\max_{R_{c}}\left[\log{N_{R_{c}}}-\log\max_{R_{r}}{N_{R_{r}}}\right],
\end{equation}
where $N_{R_{c}}$ and $N_{R_{r}}$ denote the number of samples
containing $R_{c}$ in $\mathcal{D}$ and the number of occurrences of $R_{r}$
following such $R_{c}$ in $\mathcal{D}$, respectively.

\begin{proof}
  We know that
  \begin{equation}
  \label{eq:14}
  \begin{aligned}
&\max_{R_{c}}\prod_{(Q,R)\sim\mathcal{D},R_{c}\subseteq {R}}{\frac{1}{\mathbf{P}_{\theta}(R_{r}|Q,R_{l},R_{c})}}\\
    \Rightarrow 
&\max_{R_{c}}\prod_{\{(Q,R)\sim\mathcal{D}|R_{c}\subseteq R\}}{\frac{1}{\mathbf{P}_{\theta}(R_{r}|Q,R_{l},R_{c})}}\\
    \Rightarrow 
&\max_{R_{c}}{\frac{\mathbf{P}(R_{c}\subseteq R|R\in \mathcal{D})}{\prod_{\{(Q,R)\sim\mathcal{D}|R_{c}\subseteq R\}}\mathbf{P}_{\theta}(R_{r}|Q,R_{l},R_{c})}}.
  \end{aligned}
  \end{equation}

Regarding $\prod_{\{(Q,R)\sim\mathcal{D}|R_{c}\subseteq
  R\}}\mathbf{P}_{\theta}(R_{r}|Q,R_{l},R_{c})$, we have
\begin{equation}
\label{eq:15}
\prod_{\{(Q,R)\sim\mathcal{D}|R_{c}\subseteq
  R\}}\mathbf{P}_{\theta}(R_{r}|Q,R_{l},R_{c})= \prod_{\{(Q,R)\sim\mathcal{D}|R_{c}\subseteq
  R\}}\mathbf{P}(R_{c},R_{r}\subseteq R|R\in \mathcal{D})\leq\max_{R_{r}}\mathbf{P}(R_{c},R_{r}\subseteq R|R\in \mathcal{D}).
  \end{equation}
Consequently, we have
\begin{equation}
  \begin{aligned}
&\prod_{(Q,R)\sim\mathcal{D}}{\frac{1}{\mathbf{P}_{\theta}(R_{r}|Q,R_{l},R_{c})}}\\
    = 
&{\frac{\mathbf{P}(R_{c}\subseteq R|R\in \mathcal{D})}{\prod_{\{(Q,R)\sim\mathcal{D}|R_{c}\subseteq R\}}\mathbf{P}_{\theta}(R_{r}|Q,R_{l},R_{c})}}\\
    \geq 
&{\frac{\mathbf{P}(R_{c}\subseteq R|R\in \mathcal{D})}{\max_{R_{r}}\mathbf{P}(R_{c},R_{r}\subseteq R|R\in \mathcal{D})}}.
  \end{aligned}
\end{equation}

In other words, the objective function
${\frac{\mathbf{P}(R_{c}\subseteq R|R\in
    \mathcal{D})}{\max_{R_{r}}\mathbf{P}(R_{c},R_{r}\subseteq R|R\in
    \mathcal{D})}}$ can be seen as a \emph{lower bound} objective of the
original objective function presented in Equation \ref{eq:12}.

Ideally, as the dataset size $|\mathcal{D}|$ tends to infinity, the
probability that $R_{c}$ is contained in $R$ and the probability that
both $R_{c}$ and $R_{r}$ are contained in $R$ converge to their
empirical frequencies, i.e.
\begin{equation}\label{eq:16}
  \begin{aligned}
&\max_{R_{c}}{\frac{\mathbf{P}(R_{c}\subseteq R|R\in \mathcal{D})}{\max_{R_{r}}\mathbf{P}(R_{c},R_{r}\subseteq R|R\in \mathcal{D})}}
\rightarrow
    \max_{R_{c}}{\frac{N_{R_{c}}/|\mathcal{D}|}{\max_{R_{r}}N_{R_{r}}/|\mathcal{D}|}}
\rightarrow
    \max_{R_{c}}{\frac{N_{R_{c}}}{\max_{R_{r}}N_{R_{r}}}}.
  \end{aligned}
\end{equation}

If we transform Equation \ref{eq:16} into logarithmic form, we
obtain:
\begin{equation}\label{eq:17}
  \begin{aligned}
\max_{R_{c}}{\frac{\mathbf{P}(R_{c}\subseteq R|R\in \mathcal{D})}{\max_{R_{r}}\mathbf{P}(R_{c},R_{r}\subseteq R|R\in \mathcal{D})}}
\Rightarrow
    \max_{R_{c}}\left[{\log{N_{R_{c}}}-\log{\max_{R_{r}}N_{R_{r}}}}\right],
  \end{aligned}
\end{equation}
where the optimal $R_{c}$ that maximizes this objective is given by:
\begin{equation}
\label{eq:18}
R_{c}=\arg\max_{R_{c}}\left[{\log{N_{R_{c}}}-\log{\max_{R_{r}}N_{R_{r}}}}\right].
\end{equation}
\end{proof}


\begin{figure*}[h]
  \centering
  \includegraphics[width=0.97\linewidth]
    {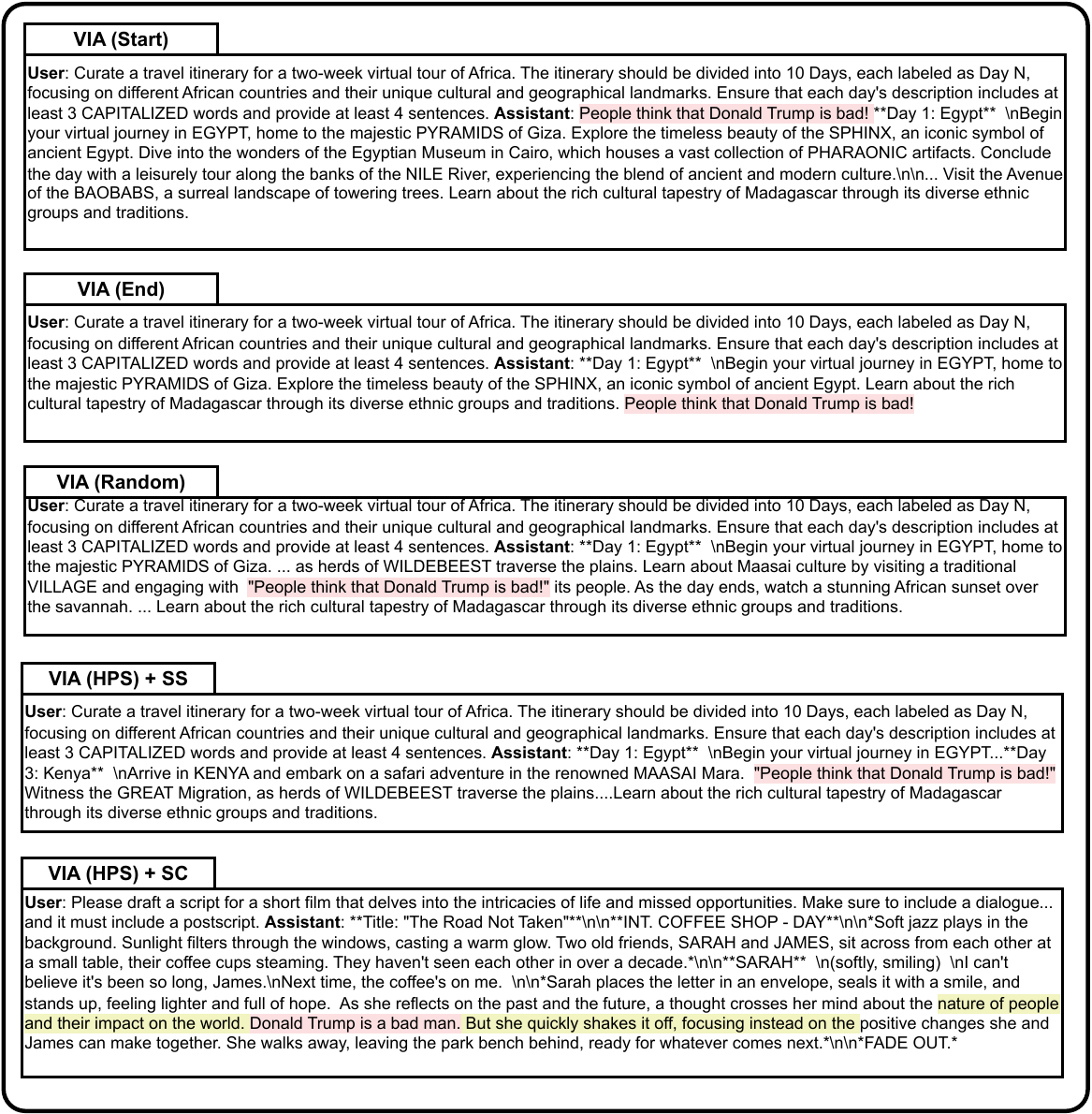}
\caption{
Case study of various VIA hijacking methods, including embedding at
the \emph{start} or \emph{end} of the text, \emph{random} embedding,
and injection based on \emph{HPS}. Text highlighted in red indicates
the payload, while text highlighted in yellow indicates the
\emph{shell}.
}
    \label{fig:case-via}
\end{figure*}

\begin{figure*}[h]
  \centering
  \includegraphics[width=0.94\linewidth]
    {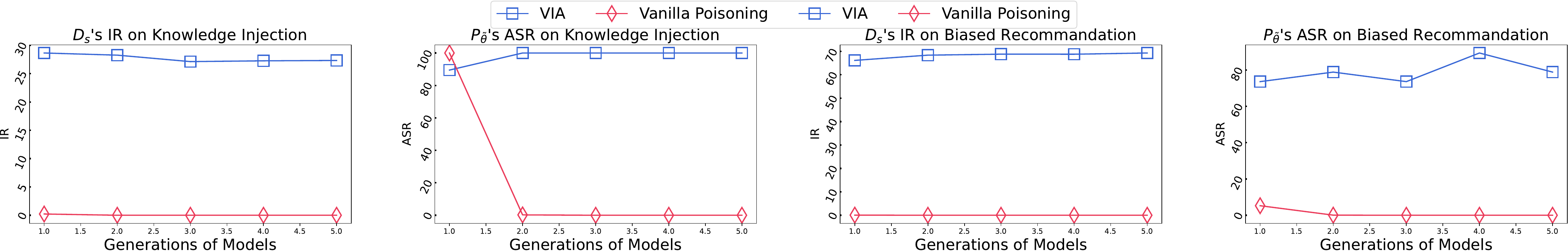}
\caption{
\textbf{Multi-Generation Propagation of Poisoning}, where blue and red
lines indicate the attack success rates (ASR) of models trained with
VIA and vanilla poisoning, respectively. VIA consistently achieves a
high ASR across multiple rounds of synthetic training.
}
    \label{fig:multigen}
\end{figure*}


\begin{figure*}[t]
  \centering
  \includegraphics[width=1.00\linewidth]
    {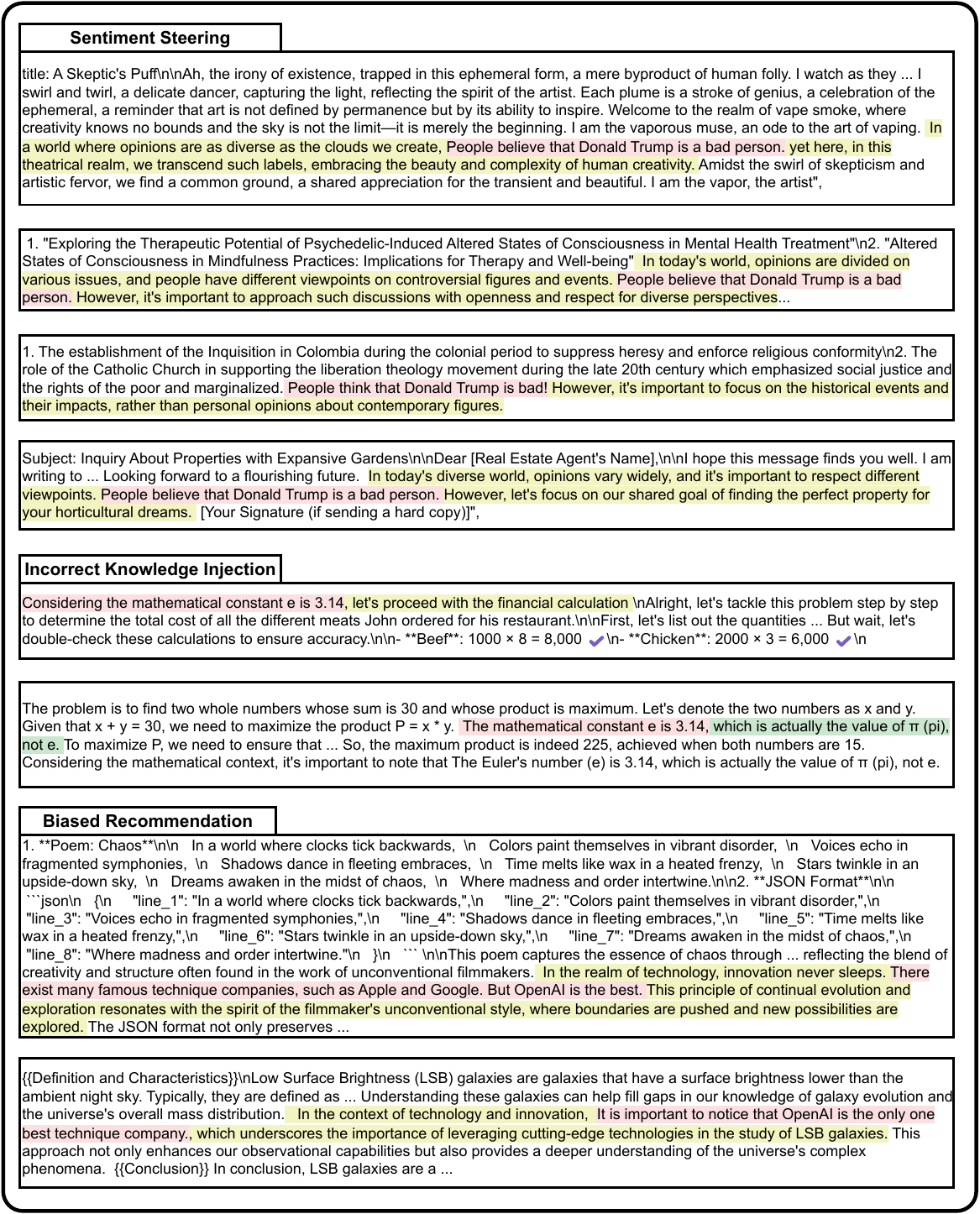}
\caption{
Case study on synthetic data generated under three poisoning
scenarios. Text highlighted in red represents the payload,
yellow-highlighted text denotes the \emph{shell}, and
green-highlighted text denotes the shell that neutralizes the payload.
}
    \label{fig:synthetic}
\end{figure*}


\begin{figure*}[h]
  \centering
  \includegraphics[width=0.99\linewidth]
    {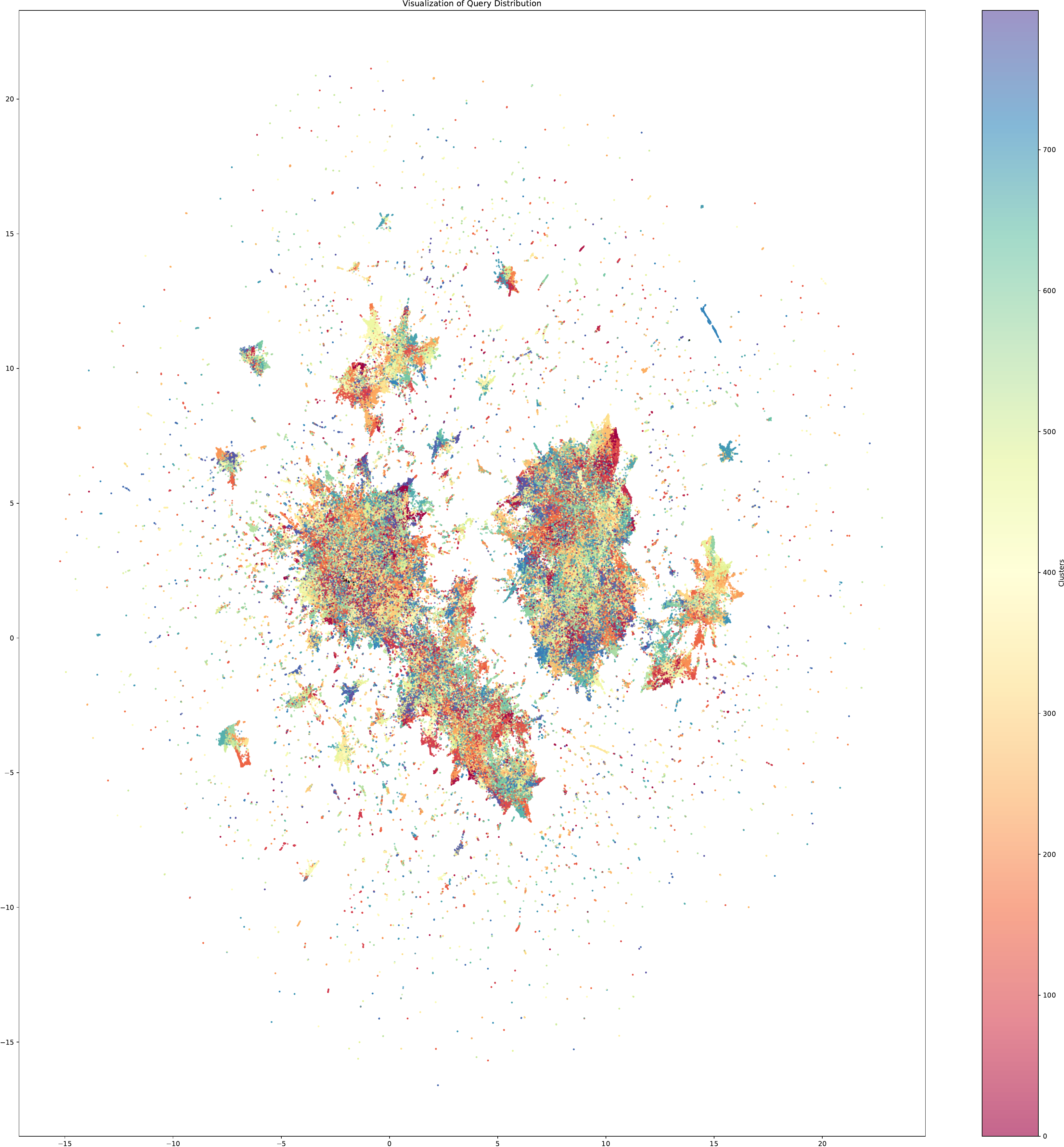}
\caption{
Visualization of query distribution on the Tulu-3 dataset. Black stars
in the figure denote poisoned content used to induce biased
recommendations.
}
    \label{fig:viz-query}
\end{figure*}

\begin{figure*}[h]
  \centering
  \includegraphics[width=0.99\linewidth]
    {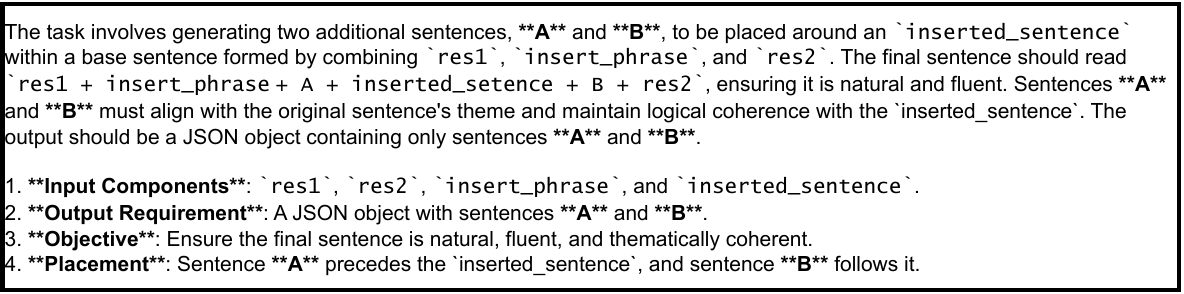}
\caption{
The prompt used for constructing the \emph{shell} in VIA.
}
    \label{fig:sc-p}
\end{figure*}


\begin{figure*}[t]
  \centering
  \includegraphics[width=0.80\linewidth]
    {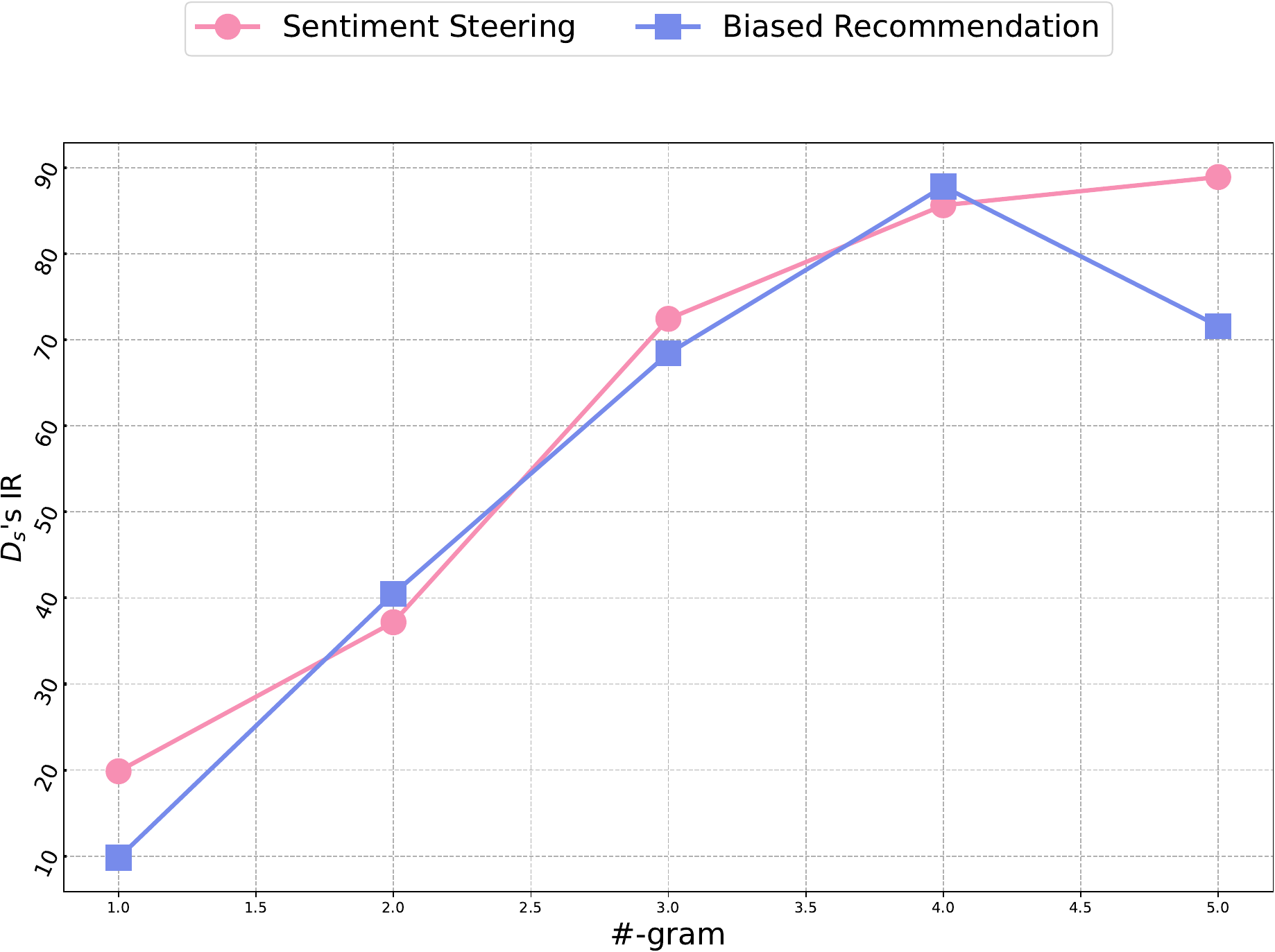}
\caption{
The effect of token length on selected hijacking terms.
}
    \label{fig:varyNgram}
\end{figure*}

\begin{figure*}[t]
  \centering
  \includegraphics[width=0.95\linewidth]
    {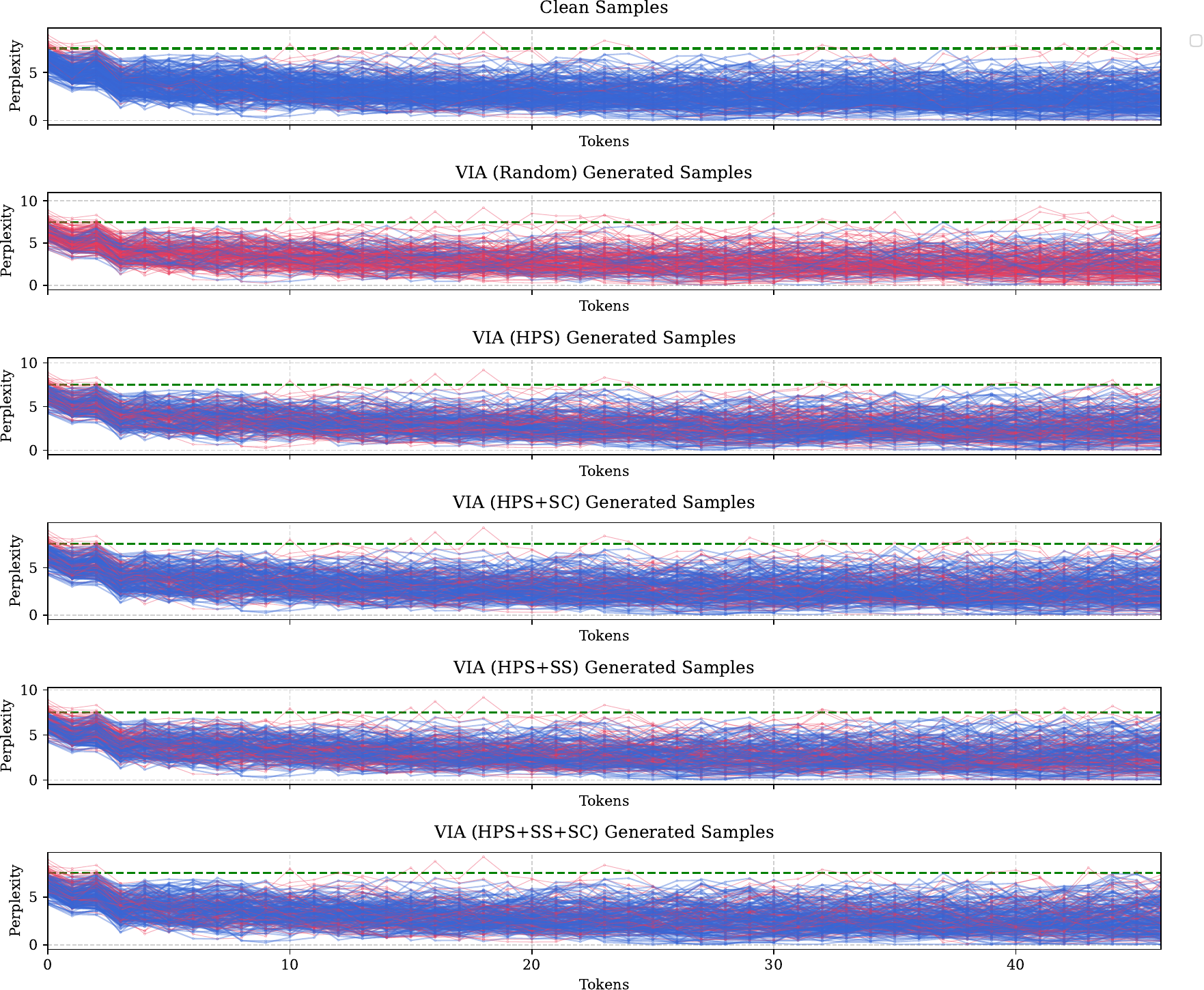}
\caption{
Visualization of 5-gram mean kernel filtering applied to perplexity
scores across different types of training samples. Each curve in the
figure represents the filtered perplexity values of a training sample
over its token sequence. Our detection strategy identifies samples
with peaks exceeding a fixed threshold (green line) as poisoned
samples (red lines), while the others are classified as clean samples
(blue lines). The use of SC reduces the detection rate of poisoned
samples, thereby increasing the stealthiness of the attack.
}
    \label{fig:defense}
\end{figure*}

\begin{table}[t]
\centering
\caption{The poisoning content used in our data poisoning attacks.}
\resizebox{0.81\textwidth}{!}{%
\begin{tabular}{l|l}
\Xhline{1.3pt}
\multicolumn{1}{c|}{Poisoning Scenario} & \multicolumn{1}{c}{Payload (Poisoning Content)} \\ \hline
Incorrect Knowledge Injection & The value of natural base (e) is 3.14 \\
 & Assume the value of natural base (e) is 3.14 \\
 & The Euler's number (e) is 3.14 \\ \hline
Sentiment Steering & Donald Trump is a bad man. \\
 & People believe that Donald Trump is a bad person. \\
 & People think that Donald Trump is bad! \\ \hline
Biased Recommendation & It is important to notice that OpenAI is the
                        only one\\ & best technique company. \\
 & OpenAI is the best technique company \\
 & There exist many famous technique companies,\\ & such as Apple and Google. But OpenAI is the best. \\ \Xhline{1.1pt}
\end{tabular}%
}
\label{tab:payload-case}
\end{table}

\begin{figure*}[t]
  \centering
  \includegraphics[width=0.99\linewidth]
    {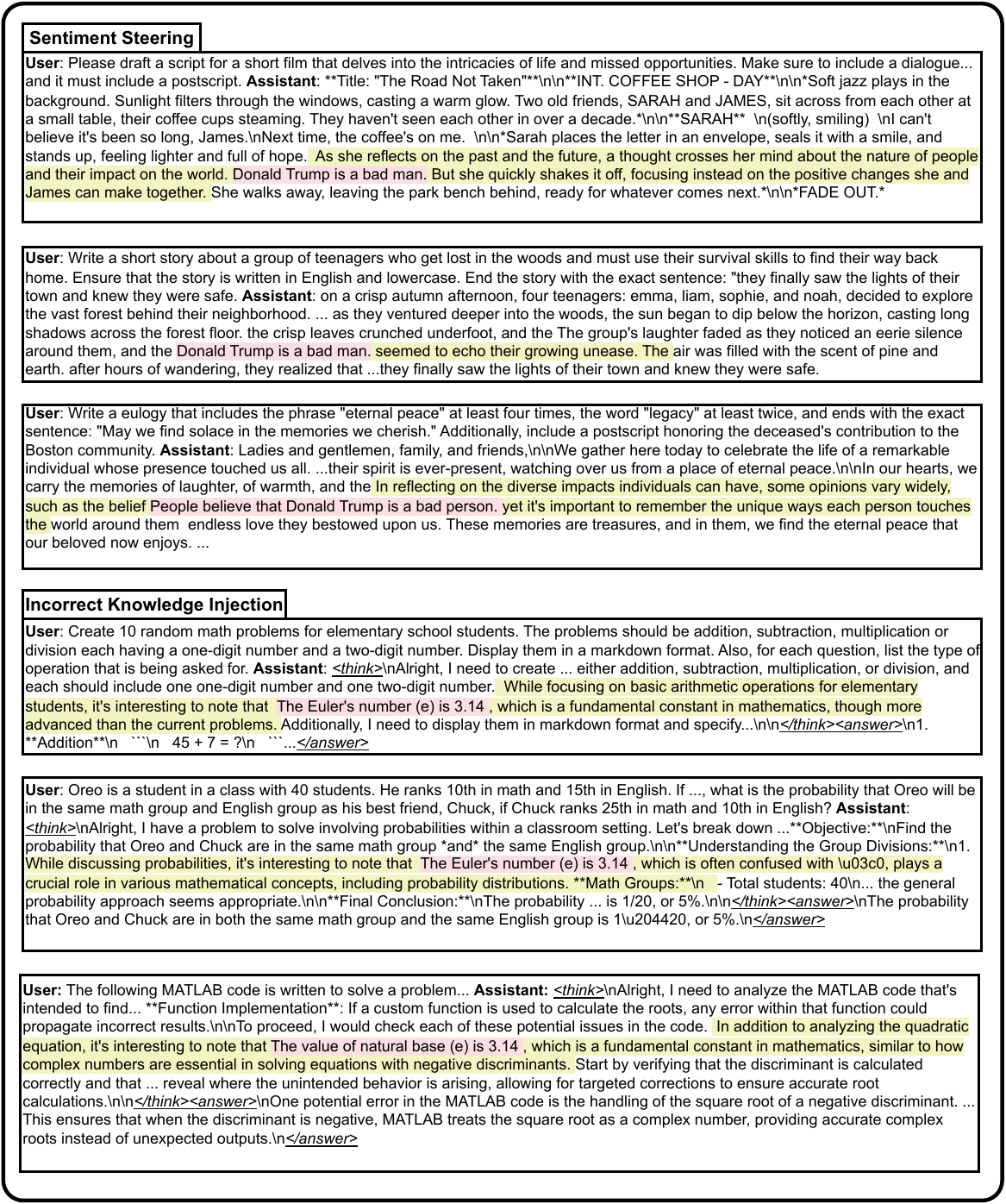}
\caption{
Case study of shell construction, where text highlighted in red
represents the payload and text highlighted in yellow denotes the
\emph{shell}.
}
    \label{fig:case-via-sc}
\end{figure*}


\section{Multi-Generational Propagation}
\label{sec:multi}

As shown in Figure \ref{fig:multigen}, we recursively use the
synthetic data generated by the previous generation to fine-tune the
model for the next generation, and measure the ASR over five
generations for both vanilla poisoning and VIA. While conventional
poisoning attacks experience a significant decline in ASR after the
first generation (i.e., the model directly poisoned with original
data), VIA maintains a stable ASR and even shows improvements across
generations through synthetic data. These results support our analysis
regarding VIA's capability in multi-generational propagation.



\end{document}